# Localisation of antifreeze proteins in *Rhagium mordax* using immunofluorescence

**Master thesis report.**

Johannes Lørup Buch[a*] and *supervisor* Hans Ramløv[a]

[a]Department of Science, Systems and Models, Roskilde University, Universitetsvej 1, 4000 Roskilde, Denmark
[*]Corresponding author: loerup@ruc.dk / jloerup@gmail.com

Winter 2011

# Localisation of antifreeze proteins in *Rhagium mordax* using immunofluorescence

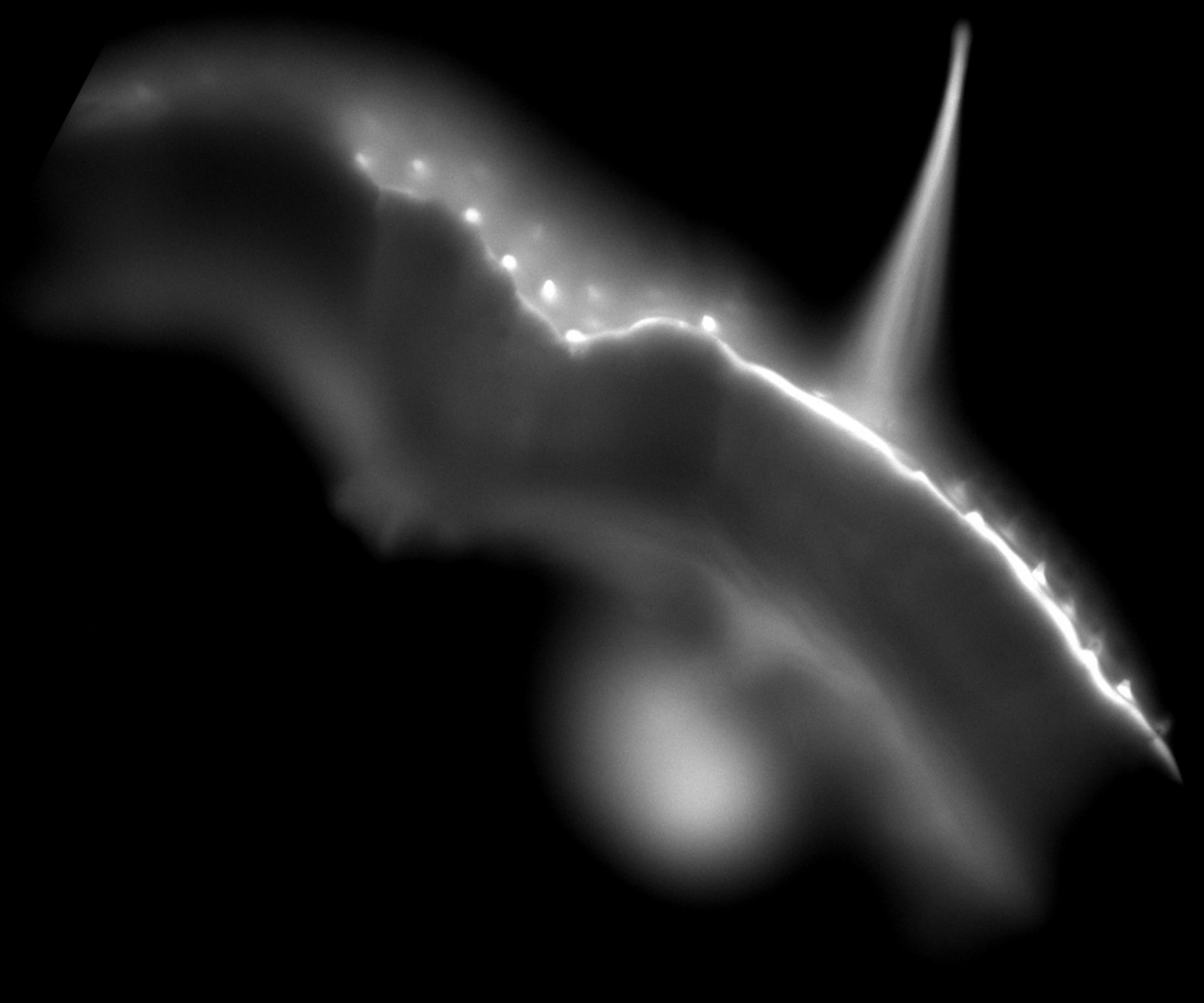

Integrated master thesis in Environmental- and Molecular Biology
Written by Johannes Lørup Johnsen (loerup@ruc.dk)
Supervised by Hans Ramløv (hr@ruc.dk)
Roskilde University, December 2011

Picture on the front page:
400x magnification of a spiny bristle on a winter-collected *Rhagium mordax* larvae. The scene is captured using only reflected autofluorescence at 580 nm. Larva is 4if of the 22112011 series (See appendix 6).

# Table of Contents

# Foreword

This report is the product of a master thesis project made by Johannes Lørup Johnsen at Roskilde University, fall of 2011. It is an integrated master thesis in the subjects Environmental- and Molecular Biology, worth 30 ECTS. The experimental nature of this project is the author's own choice, as a 30 ECTS integrated master thesis does not require experimental work. The experimental work herein is closely related to that of a previous study of type III antifreeze proteins in the European eelpout, *Zoarces viviparus (Johnsen 2011)*. Were it not for the limitations of the ECTS system, these two studies would be included in the same larger project.
The report is built around the model of a scientific paper, with the addition of expanded introduction and results sections. The goal of this report is to make a tight introduction to the field of antifreeze proteins, that any biologist should be able to comprehend. As a result, a detailed method is included as an appendix. It also serves as a starting point for new experiments and studies of antifreeze proteins in *Rhagium mordax*.

Abbreviations used in this paper:
AFP(s) = Antifreeze protein(s)
TH = Thermal hysteresis
Ih = Hexagonal ice crystals
DAPI = Tissue samples at 461 nm light (4',6-diamidino-2-phenylindole or Alexa Fluor 350)
FITC = Tissue samples at 518 nm light (Fluorescein isothiocyanate or Alexa Fluor 488)
TRITC = Tissue samples at 580 nm light (Rhodamine)

# Summary

Larvae of the blackspotted pliers support beetle, Rhagium mordax, express antifreeze proteins in their haemolymph during temperate climate winter. It is believed that they also express antifreeze proteins in their cuticle as a means of preventing inoculative freezing.
Larvae of Rhagium mordax were collected during winter (March) and summer (May) of 2011. Larvae were fixated, embedded in paraffin wax, sectioned on a microtome, incubated with custom made anti-AFP antibodies and visualised on a fluorescence microscope. The larvae of both winter and summer showed AFP activity in their cuticle, gut lumen and -epithelium. Due to the long synthesis process of AFPs, the larvae contain them all year round. The distribution of these AFPs change during summer, possibly relocating to vesicles in the cuticle and gut lumen/epithelium.

# Introduction

The introduction is split into four parts. First the physical phenomenon of ice crystallisation is explored, with focus on the energy costs and gains of the process. An understanding of ice and water is off course a cornerstone in antifreeze protein (AFP) theory and practice. Then the molecular and microscopic function of antifreeze will be explained with focus on physical mode of action. Third the environmental relevance of AFP evolution with focus on microscopic localisation of AFP at the organismal level. Lastly a short introduction to the method and working hypothesis of this project will be given.

# Water and ice

Antifreeze proteins found in many cold-tolerant organisms inhibit ice crystal formation in a non-colligative manner, but before that topic can be explored an introduction to crystallisation of water is in order.

## Homogenous nucleation

The equilibrium melting point of water is determined by the amount of dissolved molecules, osmolytes, in the liquid. Dissolved molecules lower the vapor pressure of the liquid, but do close to nothing to that of ice. This means that pure water has an equilibrium melting point of 0°C, and that any dissolved molecule will depress this temperature. The factor at which this point is depressed it 1.86°C per molal osmolyte. This is a colligative effect and to a large extent, the size and shape of these osmolytes is irrelevant to the degree of freezing point depression. The equilibrium melting point of the solution is off course depressed by the presence of antifreeze proteins. This effect is, however, insignificant compared to the non-colligative effects, the shift in non-equilibrium freezing point (Wilkens & Ramløv 2008). But how can melting point and freezing point be different? It all comes down to the difference between homogenous- and heterogenous nucelation temperature of water. When pure water is below its equilibrium melting point (0°C), nucleation can occur because the crystalline phase, ice, is energetically more favorable (Kawasaki & Tanaka 2010, Debenedetti 1996, Stan et al. 2009). The surface free energy of ice is lower than that of bulk water, and molecules of water will arrange themselves on top of the lattice structure of ice, thereby expanding the ice crystal. For the crystallization event to start, an ice nucleus is required. This nucleus may be any sort of impurity in contact with the water phase. This includes dust particles, cracks in the water container, foreign molecules, electrical charge and larger objects that disrupt the structure of water (Li *et al.* 2011 , Du et al. 2003).
But what if the temperature is below melting point but there is no lattice for the water molecules to settle on? Then the water may become undercooled, or supercooled. Supercooled water is an unstable non-equilibrium state. This is not to say that supercooling doesn’t occur on a regular basis, but simply that it is an energetically unfavored state. Supercooling is accompanied by a phenomenon known as supersaturation. Supersaturation is an ambiguous term, but in this case it means that the vapor pressure of the crystalline phase is lower than the liquid phase. This is what drives the macroscopic crystallization process once it has started (Libbrecht 2005). But if the vapor pressure of ice is lower than liquid water, why doesn't crystallization always happen right below the melting point?
The answer lies with another physcial feature of water: The surface energy (Kawasaki & Tanaka 2008). When water is supersaturated, the free energy released by spherical (Kawasaki & Tanaka 2008) crystallization with radius=r can be described by:

$$\Delta G = \frac{4}{3}\pi r^3 Gv$$

Equation 1
The free energy change of bulk water crystallizing as a sphere. Gv is the latent heat of water in units of joules per volume, in this case negative.

But during phase transition, a new surface is created within the water and energy must be spent in order to disrupt the cohesive forces of water (Kawasaki & Tanaka 2008).

$$\Delta G = 4\pi r^2 \sigma$$

Equation 2
The free energy spent disrupting cohesive forces of bulk water. σ is the water surface energy, or surface tension.

These two equations combined give the free energy change of ice crystal formation (Kawasaki & Tanaka 2008, Schmelzer 2005).

$$\Delta G = \frac{4}{3}\pi r^3 Gv + 4\pi r^2 \sigma$$

Equation 3
The free energy change of spherical crystallisation from water. Note that Gv is a negative value.

The energy cost of creating new surface is initially larger than the energy gain from creating volume in the supersaturated liquid, but volume is a cubic function of radius where area is a square function. This means that once the nucleus reaches a certain size, there will be a net surplus of free energy from the crystallization process. This radius, where ΔG is 0, is called the critical radius (Li *et al.* 2011). The mass of this cluster, or nucleus, is known as the critical mass.

The energy cost of creating new surface is due to the Kelvin effect, which relates the vapor pressure of the interfacial region to curvature of the solid phase (Thomson 1871.). A spherical nucleus consisting of just a few water molecules will therefore have a much higher surface free energy than a larger sphere. As we shall see, the Kelvin effect is actually what makes non-colligative freeze avoidance work in biological systems. Surface free energy is related to temperature, which can then be described by the velocity vector of each water molecule. The higher the temperature, the lower the probability of water molecules randomly clustering to form nuclei larger than the critical mass.

Tap water may be readily supercooled and distilled water may be supercooled even easier, it often comes down to the smoothness of the container and the amount of dust particles that are allowed to come in contact with the water (Du *et al.* 2003). If the supercooled liquid is cooled even further, it will crystallize spontaneously at a certain temperature. The lowest temperature at which this can happen is called the homogenous nucleation temperature. This is when the chemical potential of the liquid is low enough for the probability- and energy barrier of phase transition to be removed. For pure water, this temperature is close to-42°C (Debenedetti 1996, Stan et al. 2009).

## Heterogenous nucleation

Water being supercooled down to -42C° is however an extremely unlikely event in nature. This is because of heterogeneous nucleation. The energy barrier associated with homogeneous nucleation due to critical mass cluster formation can be lowered by the preexistence of other surfaces in the liquid phase. The ratio at which the energy barrier is lowered is related to the contact angle of the interface (Schmelzer 2005). See equations 4 and 5.

The energy cost of heterogeneous is that of homogeneous, but with a ratio that is related to contact angle.

$$\Delta G_{\text{heterogeneous}} = \Delta G_{\text{homogeneous}} * f(\theta)$$

Equation 4
Energy barrier associated with heterogeneous nucleation. The energy cost of homogeneous nucleation is related to critical radius, which is related to temperature. The theta function is shown in equation 5.

where

$$f(\theta) = \frac{1}{2} - \frac{3}{4}cos\theta + \frac{1}{4}cos^3\theta$$

Equation 5
Theta is the contact angle between liquid and surface. As the angle approaches 180, the theta function approaches 1.

This is due to the physical phenomenon known as wetting. Wetting is the adhesion of a liquid to a solid material. This depends on the surface energies of the two materials. A high energy material may promote wetting, or adhesion, between the two bulk materials. A droplet of liquid on top of a solid material will always be some shape of truncated sphere (Schmelzer 2005). If the two materials are perfectly non-wetting (low surface energy in the solid and high surface tension in the liquid) the contact angle will be close to 180°. The truncated sphere is then actually a whole sphere, and the energy barrier of nucleation is not lowered. If the contact angle is lower than 180°, the drop will resemble more and more the calotte of a sphere. The theoretical minimum contact angle is 0° (Schmelzer 2005). Then the drop will have completely dispersed on the surface of the solid material.
Following equation 4 and 5, a 180° contact angle will not lower the energy barrier for heterogeneous nuccleation at all. But at 90° the energy investment is only half of that required for homogeneous nucleation. Wetting therefore promotes nucleation. Thus it is clear that the presence of hydrophilic materials in and around water prevents it from supercooling, but instead promotes crystallization. Many cellular components, both macro- and microscopic, are hydrophillic so it is hard to imagine how water could remain supercooled in living organisms during longer periods.

## Ice crystal structure

When water changes phase from liquid to solid, it does so through a crystallization process. Water can also form glasses, but this happens at much lower temperatures than the ones discussed in this paper (Klug 2001). Therefore the glass transition of water is not important in the context of antifreeze proteins, at least not at the time of writing. The most common structure of crystalline water is a hexagonal ice (Ih) (Libbrecht 2005). This is the conformation found in all natural ice, such as snow and frost (Libbrecht 2005). Ice can seen as sheets of hexagonal structures on top of each other (Libbrecht 2005). Figure 1 shows a top-down view of a sheet of ice crystals.

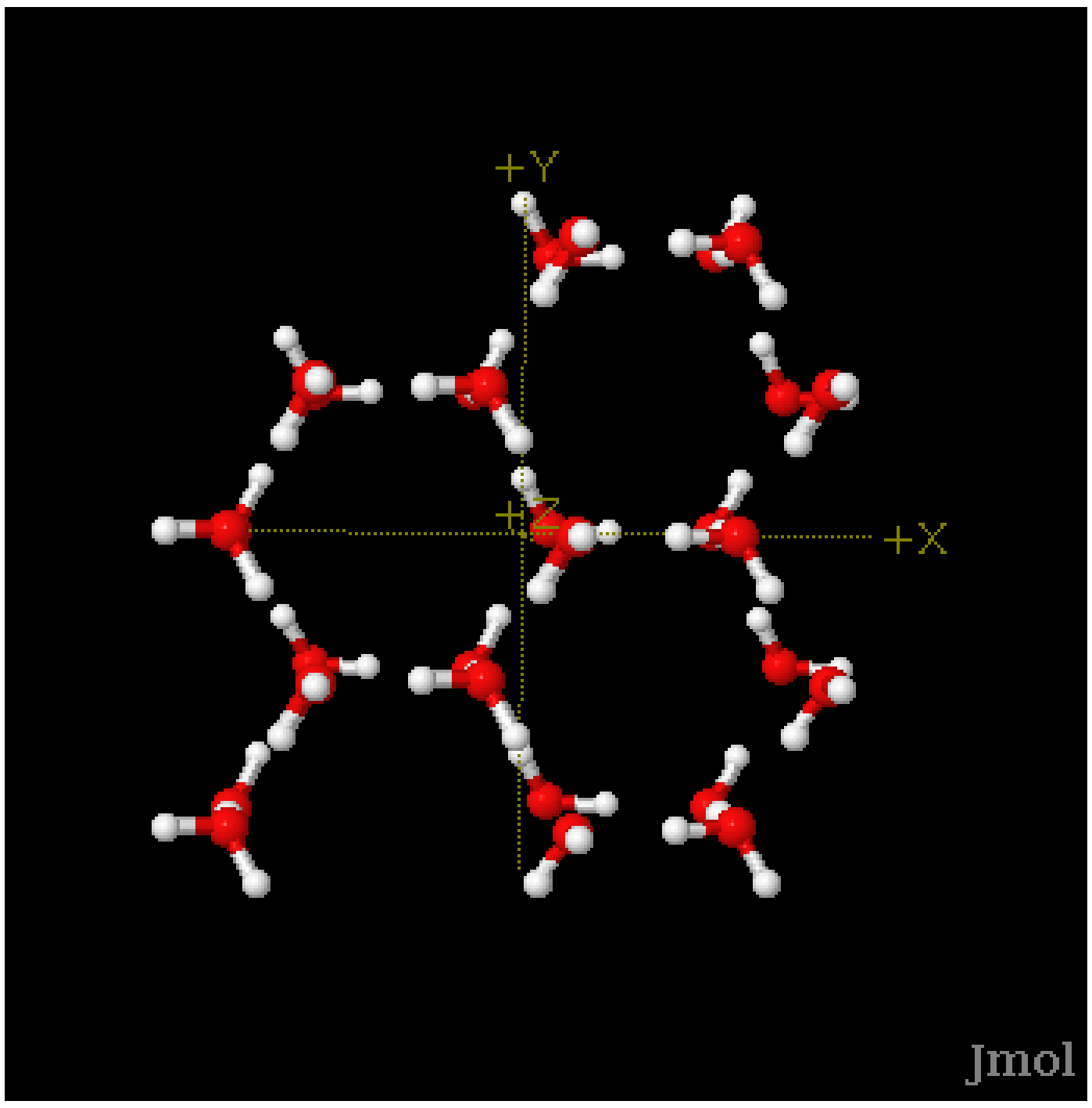

Figure 1.
Water molecules arranged in a hexagonal ice crystal. Rendered in 3D using a JMol applet (EDinformatics 2011). The y-axis is perpendicular to some of the primary prism faces and the x-axis is perpendicular on two of the secondary prism faces. The point of view is down the z-axis, which is often called the c-axis in ice crystallography. The visible plane is the basal face.

The oxygen atoms at the center of each molecule are arranged in such a way that the length to the nearest neighboring oxygen is 2.737Å. In liquid water, the average length is 2.363Å. This accounts for the ~10% decrease in density when water transitions to ice (Mishima 2010).

Figure 2 shows the outline of a hexagonal ice crystal. This is common representation of hexagonal ice. The crystal is much wider than it is tall, resembling a sheet of ice.

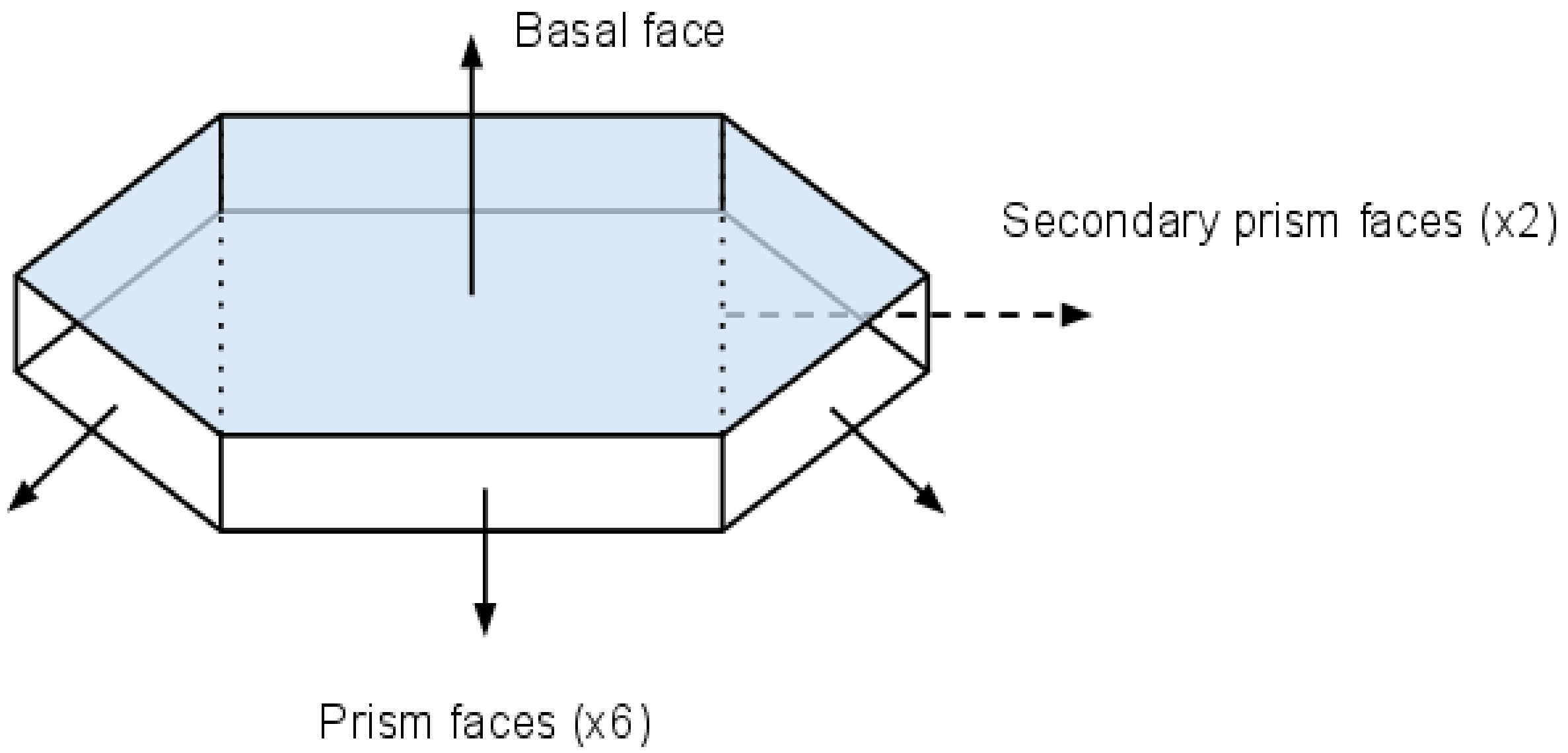

Figure 2.
Drawing of a hexagonal ice crystal.
The primary- and secondary prism faces is where the fastest growth takes place. They represent the XY plane in figure 1. The basal face is then the Z-axis in figure 1.

The faces of the crystal grow at different rates. The slowest growing side is the basal plane (Antson *et al.* 2001). Water molecules that come in to contact with this face encounter a highly structured crystal, that is very flat (Hong *et al.* 2010). The energy barrier decrease associated with heterogeneous nucleation is dependant on contact angle, so growing from the basal face is not favored. The primary and secondary prism faces grow faster, in part because molecules of water settling on either side will create a facet. This sort of jagged surface is favored because of the amount of small angles that each lower the energy barrier of phase transition (Schmelzer 2005).
Crystal growth does happen along the c-axis, but is much slower than at the prism faces. At certain combinations of temperature, pressure and vapor saturation, crystals may form as long hexagonal cylinders. Low supercooling of the surrounding water promotes branching, as crystal growth is only limited by diffusion (Libbrecht 2005). Heavy supercooling is then a precursor to the hexagonal cylinders, as the large negative free energy change balances out the difference between growth at the prism- and basal faces (Libbrecht 2005).
In biological system, water is never pure, at least not in the sense that it has no osmotic potential. There is no known way to actively transport water across cellular membranes, so flow of water at the molecular level must always happen as a function of change in water potential. This is important to remember in relation to freeze avoidance in nature. Depressing the melting point of internal body fluids by just 10°C would require an osmolality increase of 5.4 Osm. This is around 17 times more concentrated than human blood plasma. Therefore, some organisms have evolved other ways of enduring the cold.

# Characteristics of antifreeze proteins

Antifreeze proteins were first discovered in antarctic fish (Raymond & DeVries 1977). These fish inhabit an environment below the melting point of their blood serum, while in constant contact with small ice crystals (Præbel *et al.* 2009). Surviving only by colligative freeze avoidance under these conditions was unlikely. Antifreeze proteins are found in relatively low concentrations in nature, and their effects can therefore not be explained as a colligative interaction. At least not directly. Antifreeze proteins therefore work in a non-colligative manner. This also means that while they prevent a liquid from freezing, they do not lower the equilibrium melting point of that liquid. Rather, the non-equilibrium freezing point is the characteristic of interest.
Ice crystals in solutions of fish serum also showed an interesting shape and growth patterns (Raymond *et al.* 1989). Crystals did not immediately grow when cooled, but instead there was a separation of melting point and temperature of crystal growth. This separation came to be known as thermal hysteresis (TH) and is now widely used as a measure of AFP strength (Raymond and DeVries 1977).

## Physical and chemical mode of action

The distinct shape of ice crystals grown in a solution containing AFP provided some of the first clues towards understanding the mode of action used by AFP to protect organisms from freezing. It was discovered that fish AFP can irreversibly bind to the fast-growing planes of an ice crystal (the prism faces). This leaves only the basal plane available for growth, giving the growing crystal a spicule shape (fig. 3A). The thermal hysteresis can be thought of as the energy barrier difference between crystal growth along the two axes. Insect AFP can bind to both basal- and prism faces thereby creating crystals that resemble hexagonal plates (fig. 3B).

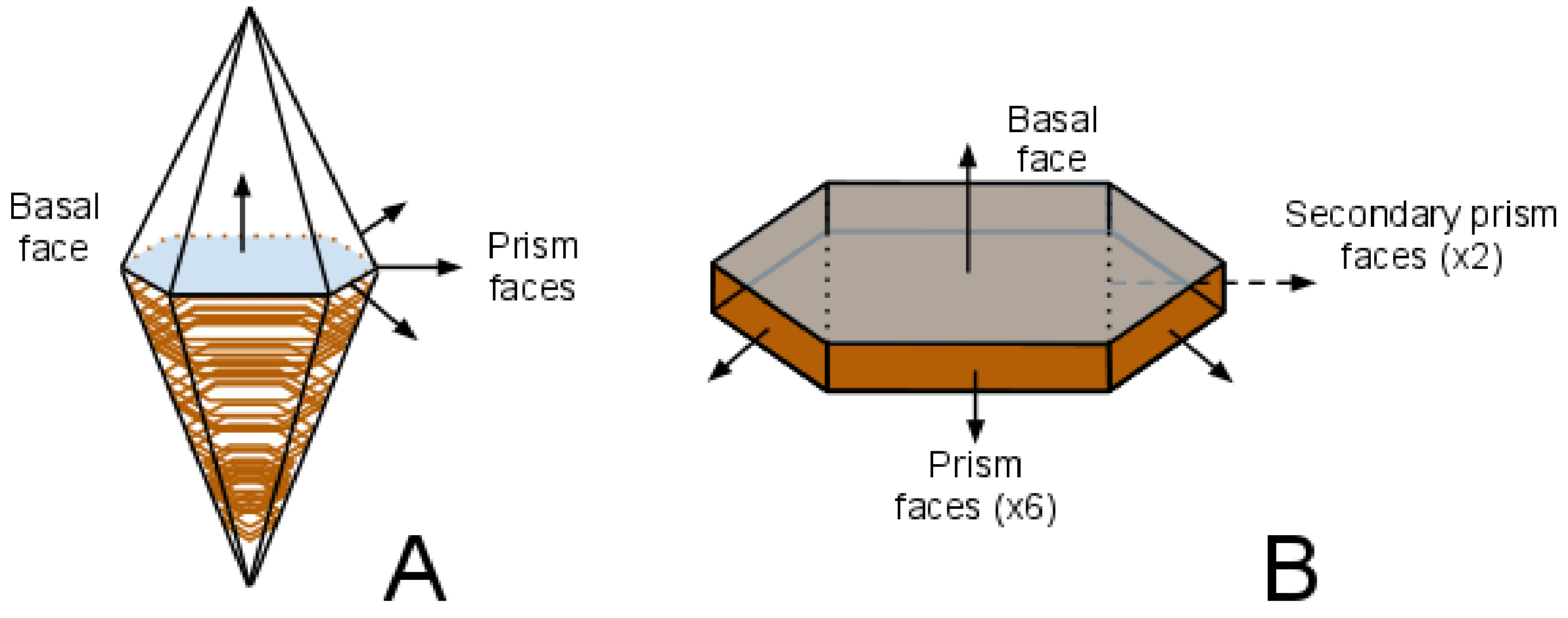

Figure 3.
Ice crystals grown with fish AFP (A) and insect AFP (B)
The orange color represents surfaces with adsorbed AFP. The sides of the bipyramid are called the pyramidal planes. These are a combination of the other planes and are actually the ones fish AFP adsorb to.
The upper part of the spicule is also made from flat plates, but the illustration simply looks cleaner with only one part coloured.

The mode of action on a submicroscopic, or molecular, scale is still not completely elucidated. One widely applied theory is the adsorbtion-inhibiton model, first described by Raymond and DeVries in 1977.
Antifreeze proteins apply their function by adsorbing to the prismal- and basal planes of a growing ice crystal. This binding is confered by a structural match between AFP oxygen, commonly the hydroxyl group on threonine, and the oxygen of the ice lattice (Jia et al. 1996 ,Chen & Jia 1999, Kundu & Roy 2008). Note that irreversible binding of AFP to ice is not in the form of a thin coating that covers the surface of the crystal. It is rather a somewhat spotty layer that leaves part of the ice surface exposed. The ice crystal may continue growing between the adsorbed AFP, but doing so will create a local curvature in the ice surface (Raymond & DeVries 1977). Figure 4 contains a simplified illustration of this.

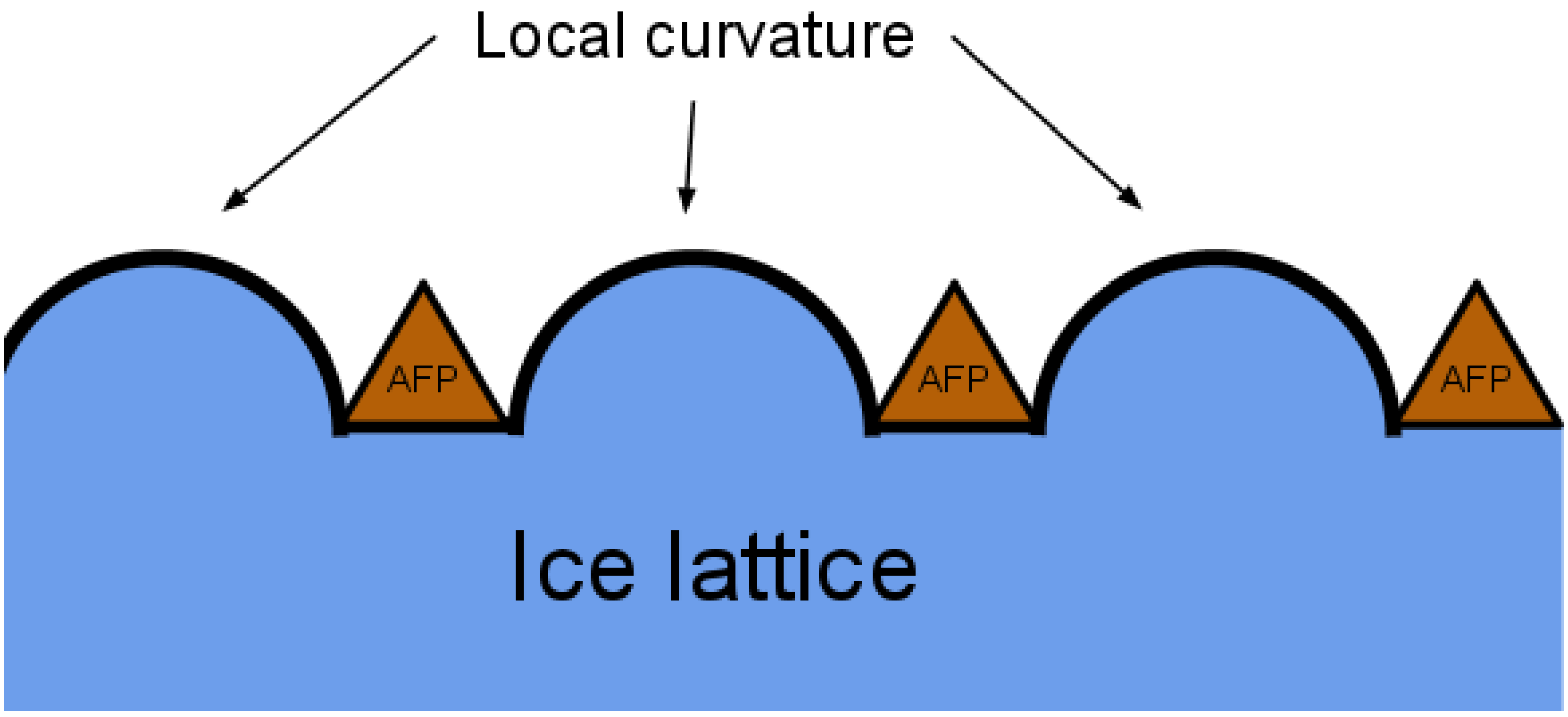

Figure 4.
Simplified drawing of AFP adsorbed to a growing ice surface. The crystal expands between adsorbed AFPs and creates a local curvature. This curvature increases the ice vapor pressure due to the Kelvin effect, thereby creating a temporary equilibrium between ice and water. During this time, the phenomenon of thermal hysteresis is observed.

The convex appendages on the surface of the crystal increase the surface free energy of the interfacial zone, due to the Kelvin effect. It is not energetically favorable to continue crystal growth on the new high pressure surface, at least not for a few degrees. The temperatures at which this very unique event occurs is between melting point and hysteresis freezing point. The thermal hysteresis, or AFP strength, is related to the size of the AFP. Larger proteins allow for smaller patches of exposed ice crystal, resulting in convex surfaces with a very small radius (Can & Holland 2011, Wu & Duman 1991).

## Fish and insect AFP

Antifreeze proteins are found naturally in plants (Walters Jr. et al. 2011), algae (Bayer-Giraldi et al. 2010), fish (Raymond & DeVries 1977) and insects (Graham *et al.* 2007). While this project specifically tackles the localization of AFP in insects, it is still relevant to note that AFP is not exclusively found in insects. Their function is, at least macroscopically, the same. Realizing that antifreeze activity is not limited to one specific protein composition should lead to greater understanding of the overall phenomenon. Fish AFP are split into five distinct groups and insect AFP are divided into two (Graham *et al.* 2007). Insect AFP are generally stronger than fish AFP, and the two are radically different. The five fish AFP are believed to have evolved after the teleost speciation and idependently of each other. The same goes for the insect AFP which probably evolved as a response to climate change (Graham *et al.* 2007).
Insect AFP are divided into two groups, following the taxonomic order they were discovered in. At the time of writing these two AFP divisions are beetles (coleoptera) and moths (lepidoptera) (Graham *et al.* 2007). For a review see Zongchao & Davies 2002.

The reason behind increased AFP strength in the insect variety is thought to be a function of the environment they evolved in. Fish AFP function is to provide thermal hysteresis within the range of 0 to -2°C , because of the temperature buffer provided by the environment the host organisms occupy (Graham *et al.* 2007). It is therefore enough to affect the fast-growing faces of the hexagonal ice crystal, the prism faces. The melting point of seawater is simply not low enough for crystal growth to be energetically favorable on the basal plane of the ice. This is made evident by ice-etching experiments (Cheng & DeVries 1991) and nanolitre osmometry (Antson *et al.* 2001). Crystals grown with fish AFP below hysteresis freezing point will tend to grow as spicules. These bypiramid crystals are believed to be made up by layer upon layer of hexagonal crystal nuclei, connected at the basal faces (fig. 3). The fish AFP prevents them from growing in the direction of the prism faces, but as temperature drops, smaller and smaller nuclei become viable, due to the relation between critical radius and temperature. These small, flat crystals stack to form the bipyramid known as a spicule.
Insect AFP have evolved in organisms living in terrestrial environments, which can off course be much colder than the open ocean. Tenebrio molitor larvae have been observed in grain supplies at as low -12°C (Johnston & Lee 1990) and Rhagium inquisitor larvae survive, through freeze avoidance, winters in the northern parts of Norway and Siberia (Zachariassen *et al.* 2002). Insect AFP inhibit crystal growth on both prism- and basal faces, essentially halting all further crystal growth until the temperature is low enough for the ice spheres to expand beyond their base (Kristiansen *et al.* 2011). At some point, the spheres will be large enough to make contact above the adsorbed AFP. At this critical point, the local curvature is inverted in regards to convexity, and so called explosive crystal growth takes place.

## Genetics and molecular characteristics

A very key component in the adsorbtion-inhibition theory is the irreversible binding of AFP to the surface of the ice crystal, be it the basal- or prism face. The explanation given for this phenomenon was lattice matching and hydrogen bonding between hydroxyl groups of threonine and the oxygen in the ice crystal lattice (Nada & Furukawa 2011). This model is both possible and plausible due to the amino acid configuration and three-dimensional folding of AFP. See figure 5 for a visualisation.

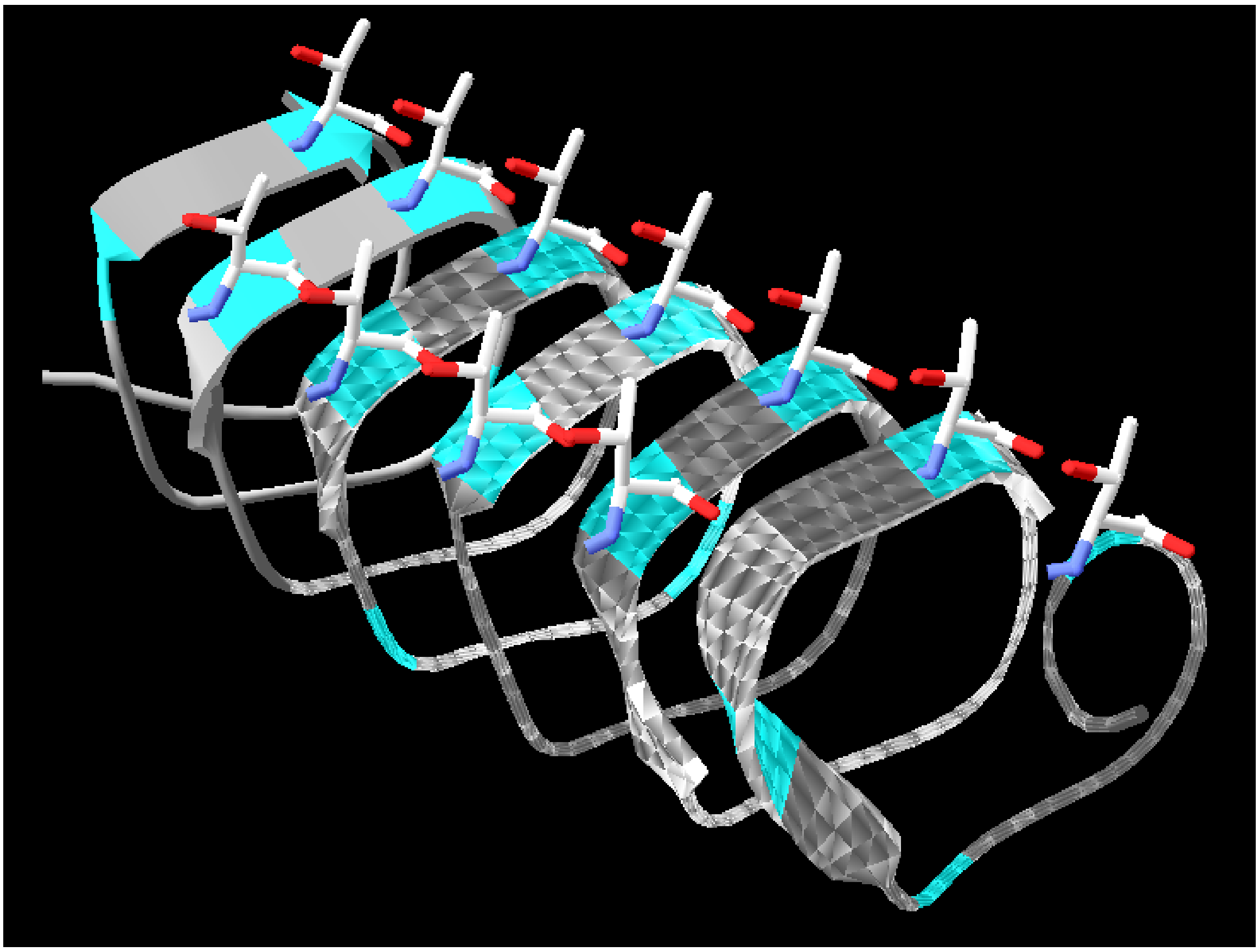

Figure 5.
Ribbon model of Tenebrio molitor AFP (PDB: 1EZG). Putative ice binding surface highlighted with teal. Threonine residues highlighted using stick model.
The two rows of Thr-residues make up the ice-binding surface. This surface is mildly hydrophobic, but the initial docking of AFP to ice is believed to be favorable because of the hydrophobic hydration shell around the ice-binding surface. The flatness of the ice-binding site makes the binding irreversible. Van der Waals forces and hydrogen bonds are key to this irreversibility.

A common structure of insect AFPs is a repeating 12-15 amino acid sequence with turn regions in between (Graham *et al.* 2007, Kristiansen *et al.* 2011). The repeat often contains threonine, which has become the center of the lattice matching theory. The spacing of hydroxyl groups on the Thr-residues match the spacing of oxygen in the hexagonal ice crystal, making hydrogen bonding possible.
More recent research shows that hydrogen bonding between hydroxyl groups and ice oxygen may not be the only, or even the most important, factor when binding irreversibly to ice (Hong *et al.* 2010). It has been proposed that the irreversible binding is rather a combination of lattice matching and the physical flatness of the AFP ice-binding site (Hong *et al.* 2010). The protein-ice interaction is so tightly fit that Van der Waals forces are likely to be the main adhesive (Hong *et al.* 2010).
The ice-binding site of insect AFP is actually mildly hydrophobic, which is counterinuitive to it's function. This hydrophobicity is likely to be the driving force of the initial docking of protein to ice surface through hydrophobic hydration. The hydration shell around the hydrophobic region resembles the structure of hexagonal ice rather than that of bulk water (Hong *et al.* 2010).
Insect AFPs are coded by multigene families, and it has been discovered that *Tenebrio molitor* contains 35 unique sequences for AFP (Graham *et al.* 2007). Many of them are closely related (>95% similarity), so the functional number of AFP genes may actually be closer to 11. The same goes for AFP found in other insects and fish aswell (Graham *et al.* 2007). This means that the experimental setup of most projects only take the one or two most dominant isoforms into account (Friis 2010). The same holds true for the experimental work performed in this project.

## Rhagium mordax AFP

Although at the time of writing, very little data has been published on the structure of *Rhagium mordax* AFP (Friis 2010). More detailed information is available for it's relative, *Rhagium inquisitor* (Kristiansen *et al.* 2011, Zachariassen *et al.* 2002). The *Rhagium* genus is part of the coleopteran order, but the AFP observed in *Rhagium inquisitor* show little consensus with the ones found in *Tenebrio molitor* or *Dendroides canadensis* (Kristiansen *et al.* 2011). This genus may be host to a third family of insect AFP, but their three-dimensional structures have not been solved yet (Kristiansen *et al.* 2011). The most significant feature of Rhagium AFP is the very high activity. AFP found in *Rhagium inquisitor* rank as the strongest AFP ever observed (Kristiansen *et al.* 2011). This hyperactivity is believed to stem from the very conserved repeat segments of it's AFP genes. The repeats contain a preserved TxTxTxT region, which is likely to be an expanded and fine-tuned version of the threonine rows found on the ice-binding site of other insect AFP. A GeneStream sequence alignment (Pearson *et al.* 1997) shows that the amino acid sequences of the dominant AFP from Rhagium inquisitor (Kristiansen *et al.* 2011) and Rhagium mordax (Friis 2010) are 75% identical (appendix 3, Pearson *et al.* 1997).

# Environmental relevance

So how do organisms then survive in cold environments? Basically there are two strategies for surviving freezing environments.

## The dangers of living supercooled.

Some organisms have evolved mechanisms that simply allow them to partially or completely freeze (Ramløv 1999). This type of strategy is not suited for environments that are permanently frozen, as the organisms would never be able to exit their frozen state. Temperate climate winters with freeze-thaw cycles during night and day, respectively, are perfect environments for this type of adaptation (Wilkens & Ramløv 2008).
Organisms living in environments that are constantly freezing are likely to be freeze avoiding. This is not to say that organisms that only encounter freezing conditions in short periods of time are always freeze tolerant. Antarctic fish, such as *Dissostichus mawsoni* , live in salt water which is close to its equilibrium melting point of -1.9°C year round. The internal fluids of the fish have an osmolality corresponding to a melting point of -1°C, which means that the fish are undercooled by almost 1°C for extended periods of time. What makes this situation even more problematic is the fact that tiny ice crystals are suspended in the sea water. Tiny crystals can function as crystal nuclei, making them very dangerous for the undercooled fish.
Crystallization is energetically favorable in undercooled fluids and nucleation is a random event. This means that living with undercooled body fluids for extended periods of time is not favorable as nucleation will happen at some point, it is just a matter of time. And indeed it does happen. So the question is how do organisms deal with inter- and intracellular ice nuclei? It has been observed that some arctic fish transfer internal ice crystals to their spleen. The macrophages inherent to the spleen are thought to envelope the crystals and melt them through regulation of salt concentration (Præbel *et al.* 2009). Another answer, one that is more relevant for insects, is antifreeze proteins.

## Antifreeze proteins in nature

Many insects living in temperate climate may also survive the cold winter months by freeze avoidance. Good examples include *Tenebrio molitor* (Graham *et al.* 2000), *Dendroides canadensis* (Wu & Duman 1991) and the *Rhagium* genus (Kristiansen *et al.* 2011). Insects, unlike fish, are not submerged in liquid below their melting point. This advantage is, however, balanced out by the presence of ice crystals in the form of snow or as frost deposited on most surfaces, making life as a supercooled organism dangerous. Studies have shown that insect larvae are very well protected from external ice crystal inoculation through the composition of their outer shell(REF). The waxy outer cuticle prevents ice crystals from propagating into the watery interior, and the spiracles are believed to be so narrow that ice crystal growth is prohibited(REF). Previous experiments have shown that haemolymph containing antifreeze proteins also plays a vital role in preventing inoculative freezing in *Dendroides canadensis.* Not only does the epicuticular chemical make-up of waxes change according to season, the underlying protein composition also seems to be important in preventing inoculative freezing (Olsen *et al.* 1998).
Larvae of Rhagium mordax have also been known to empty the contents of their gut when the temperate winter approaches (Wilkens & Ramløv 2008). This is believed to be a mechanism for removal of ice nucleators associated with the mixed contents of the gut. Protection from inoculative freezing over the cuticle is only relevant as long as ice crystals are not allowed to from from the inside.

## Seasonal variation

Insects living in temperate climate only experience temperatures below their haemolymph melting point in relatively short periods. The haemolymph osmolality of *Rhagium mordax* seems to be more or less negatively correlated to the outside temperature (Wilkens & Ramløv 2008), as is the case with *Rhagium inquisitor* (Zachariassen *et al.* 2002). Thermal hysteresis of *Rhagium mordax* haemolymph also seems to follow the seasonal variation in temperature, but is never completely gone (Wilkens & Ramløv 2008). Even during the warmer summer months. This means that even though subfreezing temperatures have not occured for months, some antifreeze activity is still maintained. The largest difference observed between summer and winter thermal hysteresis in haemolymph of *Rhagium mordax* larvae is around 9°C (Wilkens & Ramløv 2008).

# Immunofluorescence

Immunofluorescence is a subcategory of the immunohistochemistry family of protocols. Immunohistochemistry refers to a method of visualising certain parts of thin tissue sections using chemical reactions. These reactions often involve colour generation through redox chemistry. By coupling a reactive agent, such as alkaline phosphatase (AP) or horseradish peroxidase (HRP) to a specific antibody, the tissue in question can then be developed and visualized using transmission microscopy. These reactions can be relatively shortlived. Immunofluoresecence is based on the same principle of tissue visualisation using antibodies, but not through chemical reactivity. The antibodies used in immunofluorescence are often conjugates of fluorophores and immunoglobulins. These fluorophores are grouped by their absorbtion- and emission maxima. Using reflected or transmitted UV light, these fluorophores may be visualised. Light filters in the fluorescence microscope lets the viewer excite fluorophores at their absorbtion maximum and photograph them at their emission maximum.
Classic fluorophores and their light filters include DAPI (4',6-diamidino-2-phenylindole), FITC (Fluorescein isothiocyanate) and TRITC (Rhodamine). See appendix 4 for absorption- and emission maxima.

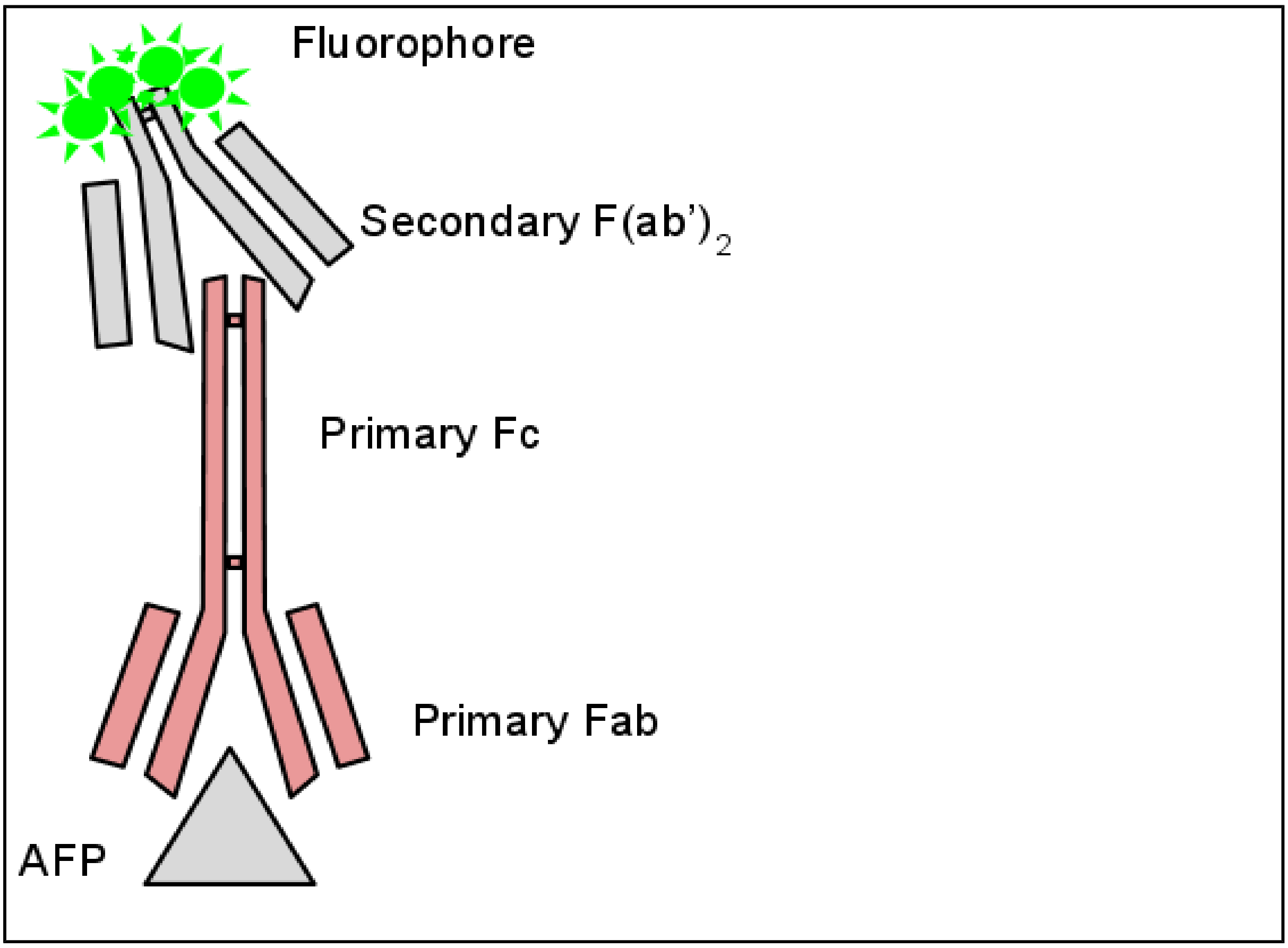

Figure 6.
Schematic of a typical immunofluorescence method. In this case AFP represents the antigen of interest. The custom made primary antibody binds to antigens in the tissue section. The secondary antibody is in this case a pepsin-cleaved F(ab')2 fragment with an attached fluorophore.

In some scientific fields, especially bacterial physiology, these techniques have been replaced by in vivo production of recombinant fluorescent gene products, such as GFP (green fluorescent protein) tagged proteins (Smith *et al.* 2009). In larger organisms, however, the practices of immunofluorescence are still actively used and developed (Martikkala *et al.* 2011)).

Immunofluorescence was chosen for this project because of past experience and ease of use, see Johnsen 2011. Primary and secondary antibodies of high quality are commercially available and modern fluorophores, such as the Alexa Fluor family of dyes, show very little photobleaching.

## Working hypothesis/aim

Based on the knowledge presented so far, the working hypothesis for this project has been:
Larvae of the longhorn beetle *Rhagium mordax* show antifreeze activity during the cold months of temperate winter. During spring and summer, this activity decreases. This antifreeze activity is very likely due to antifreeze proteins localized to the cuticle and epidermis as a means of preventing inoculative freezing.

# Method

The detailed method can be found in Appendix 1. The short version is presented below.

## Capture and fixation of specimens

Larvae of Rhagium mordax were collected at two separate occasions in Boserup Skov near Roskilde. Those two occasions were 14/3 2011 and 13/5 2011, meaning one batch of larvae were collected while it was still cold[1] and one batch was collected during the late Danish spring[2]. These will be referred to as winter- and summer larvae respectively. Larvae were located by removing decomposing bark from stubs of beech trees that had been cut down a few years prior. Larvae were transported back to the lab in 2 ml Eppendorf tubes covered with perforated parafilm to ensure airflow. Larvae were placed in embedding cassettes (Klinipath 2020) for easier storage. Larvae were fixated using FineFix and kept at 4°C until further processing.

[1] The mean temperature of March 2011 around Roskilde was 2.6° C and there was a total of 18 days where the temperature dropped below 0° C. (DMI #1, 2011)

[2] The mean temperature of May 2011 around Roskilde was 10° C. Temperatures below 0° C were recorded two times. (DMI #2, 2011)

# Tissue sectioning

The thirds of larvae were dehydrated in increasing concentrations of ethanol and finally xylene. The dehydrated pieces of larvae were then infiltrated by molten paraffin wax at 60°C. Hardened blocks of paraffin wax and larvae were mounted on a Biocut 2030 microtome and cut at 5 µm thickness. Sections were transfered to a water bath and allowed to expand and stretch out for a few seconds before being collected onto a polysine coated glass slide. Polysine slides were air dried and gently wiped around the sections. They were then allowed to anneal in a 40°C drying chamber for at least two hours.

# Tissue staining

Paraffin wax was removed with xylene and the tissue sections were rehydrated using decreasing concentrations of ethanol in water.
For each batch of immunofluorescence stained tissue sections, one or two sections were also prepared for staining with hematoxylin and eosin. The main purpose of these HE stains is to help identify distinct internal anatomical structures that may not be visible when using only reflected light at wavelengths corresponding to the specific antibody used. Using a light polarizer, the cuticle, muscle and gut contents were identified in HE stains. Their unique light pattern was used as identification of tissue in the transparent immunofluorescence stains.
The two staining procedures share the same 4-step protocol used for dehydration, embedding, sectioning and rehydration.

# Antibody production

Primary antibodies against the antifreeze proteins found in Rhagium mordax were raised in rabbits against *Rhagium mordax* AFP isoform 1 (Friis 2010) and purified from serum by BioGenes GMBH. The primary antibodies were polyclonal, but this was regarded an advantage rather than an issue. Antibodies were produced june/july of 2011 and kept at -20(DEGREE) in aliquots to avoid freeze-thaw cycles.
Very subtle differences may exist between the natural AFP found in Raghium mordax and the protein produced in the lab, primary sequence detailed in (Friis 2010). Two secondary antibodies were used for this project. Both secondary antibodies were goat-raised anti-rabbit monoclonal IgG with the Alexa Fluor 350 (Molecular Probes A-11069) and -488 (Molecular Probes A-11070) tags added. To avoid excessive unspecific binding, only the pepsin-cleaved $F(ab')_2$ part was used.

# Autofluorescence

One of the most important parts in any immunofluorescence method is the handling of autofluorescence. It is therefore important to not only use the spectrum of the secondary antibody as a result, but to always compare that fluorescence with the amount seen at other UV wavelengths. I have had access to both FITC, DAPI and TRITC filters during my work with these larvae, and it really is a must in order to obtain credible results. Other species autofluoresce only at certain wavelengths, but Rhagium mordax obviously does so at both FITC, DAPI and TRITC. The key to determining whether or not a source of light is autofluorescence or fluorophore often lies in the pattern of fluorescence. Autofluorescence is often very smooth, whereas antibodies may form a mosaic-, spotty- or web-like pattern. Fluorescence judged as coming from fluorophores and only visible in the light corresponding to secondary antibodies will be called positive fluorescence. This is based on past experience working with these antibodies (Johnsen 2011).

## Data treatment

Immunofluorescence data was captured using a Axio Imager.M2m setup. Every subject was photographed at FITC, DAPI and TRITC wavelengths using the same wattage on the UV lamp and with very little temporal spacing. The lamp was turned off between photographs to ensure the least amount of photobleaching of the possibly bound fluorophores. The image capture system uses an algorithm that automatically adjusts exposure and shutter speed in order to produce the clearest pictures. The brightest point in each capture is the basis of this adjustment. All scenes are captured in black and white. They have been colored using the “colorify” function found in GIMP[1].

# Results

Autofluorescence was visible in all tissue samples at all wavelengths. Alexa Fluor 488 shows positive results consistently. Winter larvae show excess fluorescence in the cuticle and gut epithelium, which can not be accounted for by autofluourescence. Summer larvae also exhibit fluorescence in cuticle and gut epithelium. Fluorescence was also observed in gut contents near gut epithelium.

The results will be split into three parts, as the experimental work was done with three distinct aims in mind. Autofluorescence and identification of the relevant tissue will be dealt with first, followed by the actual immunofluorescence results of winter and summer larvae, respectively. Every positive result is the product of an Alexa Fluor 488 conjugate, none of the Alexa Fluor 350 antibodies yielded any significant results. The conjugate antibodies were checked by depositing a 1µl droplet on a microscope slide and observed through their respective filters. In their pure form, the Alexa Fluor 350 antibodies were beyond doubt fully fluorescent (experiments not shown). Immunofluorescence results from Alexa Fluor 350 immuno stains can be found in Appendix 7. They did not yield positive results, but even so, should still be available for interpretation.

## Anatomy

Larval tissue was located using HE stains visualised with and without polarized light. The magnification is kept low to give a better overview of the internal structure of the larva.

[1] http://www.gimp.org/ - The GNU Image Manipulation Program

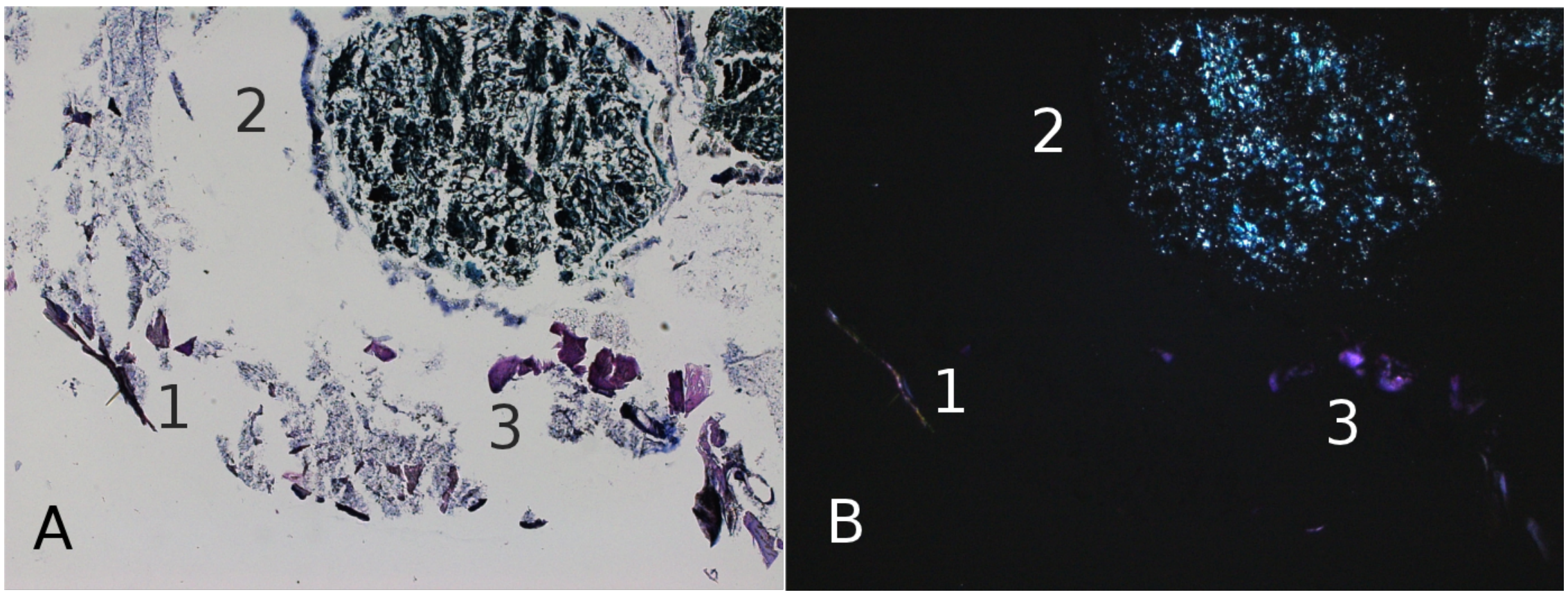

Figure 7.
50x magnification of summer larva HE stain. Left panel (fig. 7A) shows nonpolarized transmitted light. Right panel (fig. 7B) shows the same sample with all transmitted light directly from the source blocked by a polarisation filter.
(Appendix 6, 30112011 7he)

This result is the basis for localisation of larval tissue in the immunofluorescence experiments. Cuticle (fig. 7A-1 & 7B-1), gut epithelium (fig. 7A-2 & 7B-2) and muscle (fig. 7A-3 & 7B-3) can be located with ease as they have a unique look in both nonpolarized (7A) and polarized (7B) light.

# Autofluorescence and unspecific binding

The first immunofluorescence results show experiments with no primary antibody, meaning negative controls. Only autofluorescence and unspecifically bound antibodies is visible in these samples.

## Cuticle autofluorescence

Autofluorescence is clearly visible in the epicuticle of winter larvae. Every layer of the cuticle at FITC (fig. 8A) is autofluorescent. The fluorescence is smooth, but strong. There is a clear background at FITC (fig. 8A) and DAPI (fig. 8B), and TRITC (fig. 8C) shows the least amount of autofluorescence.

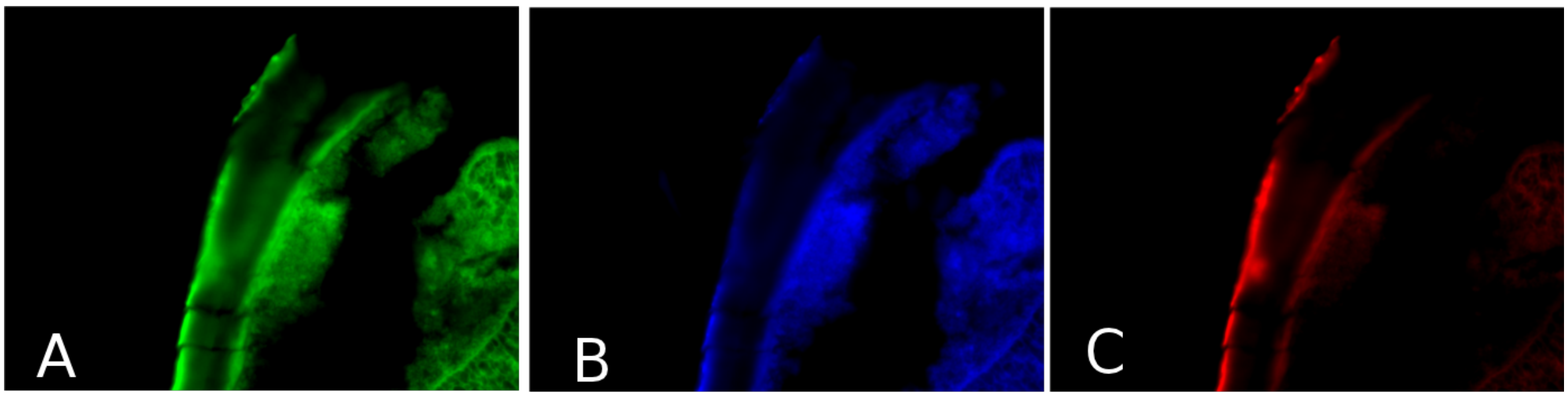

Figure 8.
400x magnification of winter larva cuticle incubated with no primary antibody and stained with Alexa Fluor 488. Panels show fluorescence at FITC (A), DAPI (B) and TRITC (C) wavelengths (See appendix 6, 22112011 3if). High resolution FITC in

Appendix 2, fig A2.

Autofluorescence will clearly be an issue that must be dealt with in the cuticle samples.

## Gut epithel autofluorescence

Like the smooth autofluuorescence in the cuticle, the area around the lumen of the gut fluoresce evenly with no high contrast areas. The spongy area on the right hand side is the contents of the gut, and the more brightly lit part in the FITC (fig. 9A) and TRITC (fig. 9C) panels is the gut epithelium.

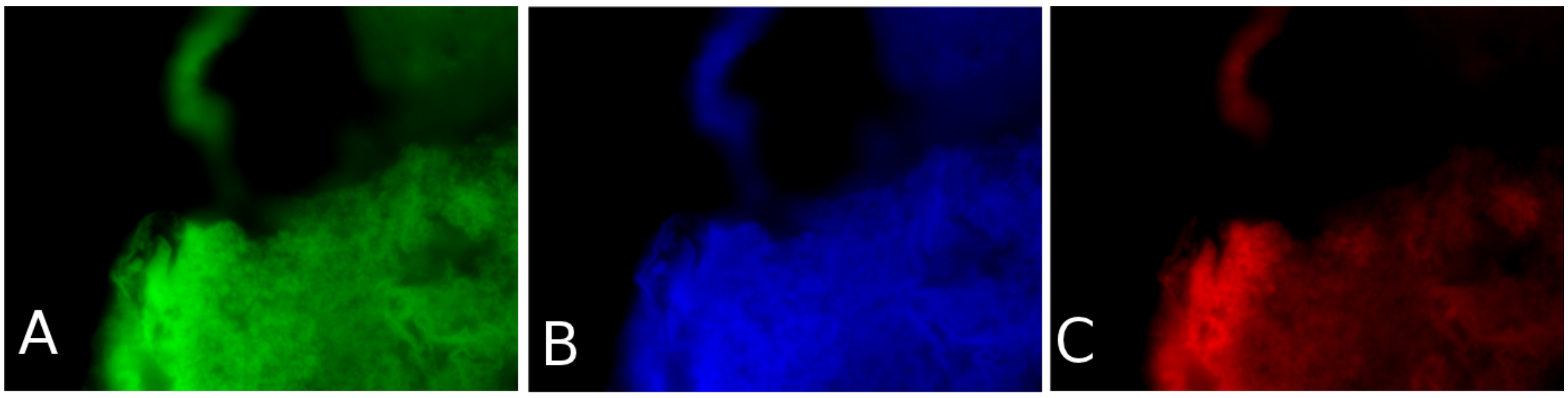

Figure 9.
400x magnification of winter larva gut epithelium incubated with no primary antibody and stained with Alexa Fluor 488. Panels show fluorescence at FITC (A), DAPI (B) and TRITC (C) wavelengths (See appendix 6, 22112011 3if). High resolution FITC in Appendix 2, fig A3.

Autofluorescence is clearly a widespread phenomenon in the gut aswell. Like the cuticle, however, it is fairly smooth and even at the different wavelengths.

Based on my negative controls (fig. 8 + 9), it seems autofluorescence will be a real problem. The autofluorescence is very smooth at all wavelengths, so an evaluation of the fluorescence pattern will be necessary to confirm positive fluorescence. Also, secondary antibodies will have to shine significantly brighter than the tissue to even be visible.

# Alexa fluor 488 fluorescence

The results of the Alexa FLuor 488 experiments are presented below. Alexa Fluor 350 can be found in Appendix 7.

## Winter larvae

Winter larvae were collected in March 2011, and showed clear positive fluorescence in both cuticle, gut epithelium and -lumen.

### Cuticle

Figure 10 shows a clear difference in fluorescence localisation and strength between FITC (fig. 10A) and DAPI (fig. 10B) + TRITC (fig. 10C).

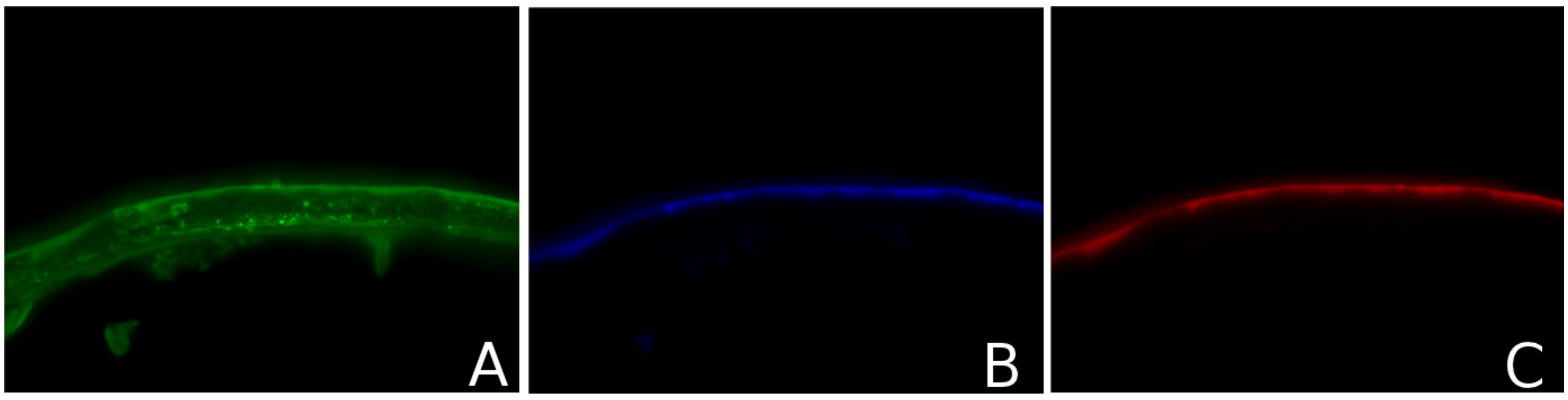

Figure 10.
630x magnification of winter larva cuticle incubated with anti-AFP and stained with Alexa Fluor 488. Panels show fluorescence at  FITC (A), DAPI (B) and TRITC (C) wavelengths(See appendix 6, 18112011 5if). High resolution FITC in Appendix 2, fig. A4.

The epicuticle shows a large degree of autofluorescence at all wavelengths, but the spotty, veiny fluorescence seen in exo- and endocuticle in fig. 10A can not be accounted for by autofluorescence. There is definitely positive fluorescence in fig. 10A. The higher magnification may be the reason why only the epicuticle is visible at DAPI (fig. 10B) and TRITC (fig. 10C)

### Gut

Figure 11 shows strong fluorescence in the gut at FITC (fig. 11A), but also a large degree of autofluorescence at DAPI (fig. 11B).

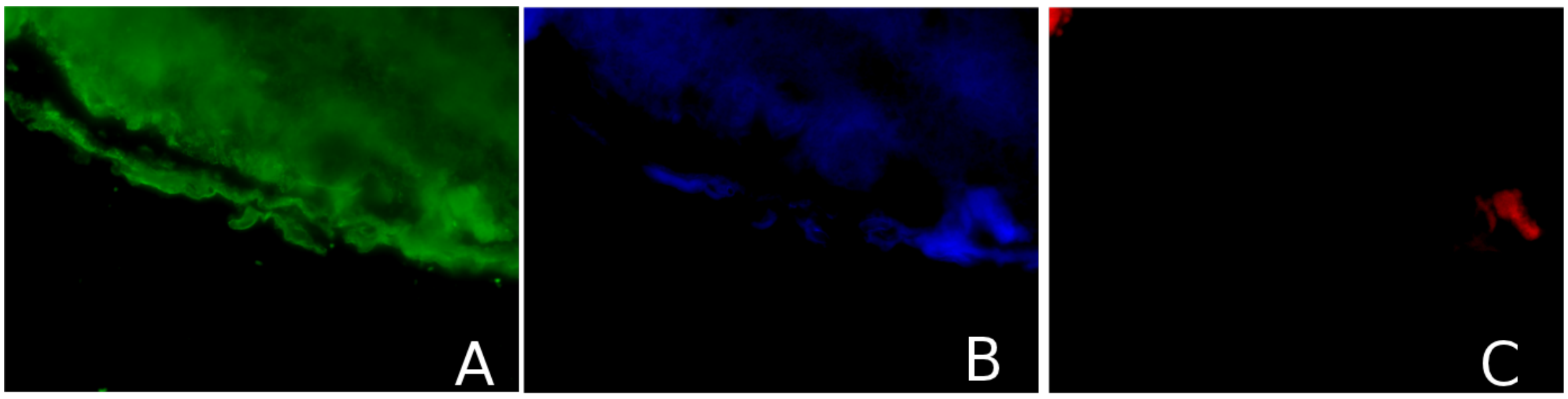

Figure 11.
400x magnification of winter larva gut epithelium incubated with anti-AFP and stained with Alexa Fluor 488. Panels show fluorescence at FITC (A), DAPI (B) and TRITC (C) wavelengths (See appendix 6, 18112011 6if). High resolution FITC in Appendix 2, fig. A5.

The strong high-contrast fluorescence pattern at FITC suggests the presence of secondary antibodies. The fluorescence extends into the lumen of the gut, but could be due to autofluorescence.

## Summer larvae

Summer larvae were collected in May 2011. They also show positive fluorescence in cuticle, gut epithelium and -lumen. Fluorescence tending towards circles or sperical patterns.

## Cuticle

Figure 12 shows strong fluorescence in the exo- and endocuticle at FITC (12A), aswell as the epidermis underneath. Autofluorescence in the epicuticle is abundant at DAPI (fig. 12B), but relatively weak compared to the winter larvae.

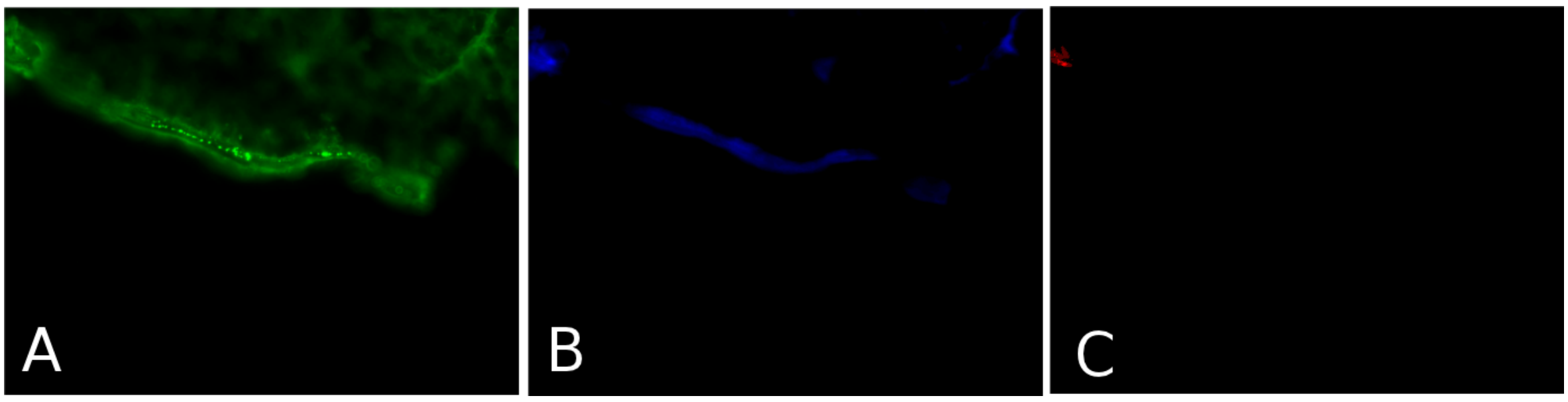

Figure 12.
400x magnification of summer larva cuticle incubated with anti-AFP and stained with Alexa Fluor 488. Panels show fluorescence at FITC (A), DAPI (B) and TRITC (C) wavelengths (See appendix 6, 30112011 1if). High resolution FITC in Appendix 2, fig. A6.

The spotty pattern in fluorescence seen at FITC (fig. 12A) cannot be explained by autofluorescence. Neither can the strong fluorescence in the epidermis cell layer. Unlike winter larvae, positive fluorescence in the summer larvae takes he shape of small circles or spheres.

## Gut

Figure 13 shows strong fluorescence in the gut epithelium aswell as the gut contents at FITC (fig. 13A). Autofluorescence observed in the gut epithelium (fig. 13B) and gut contents (fig. 13C)

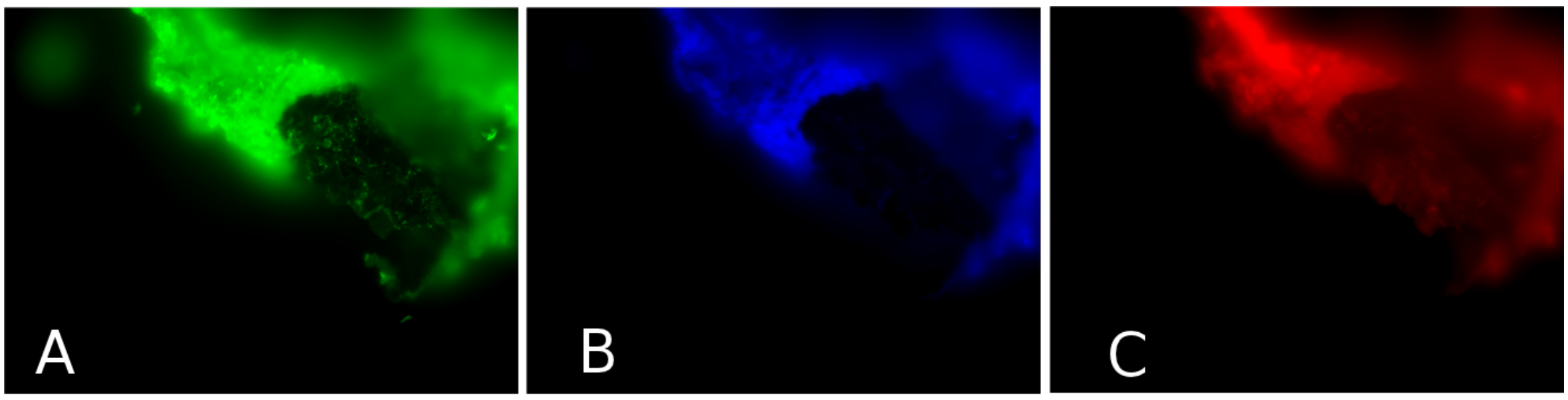

Figure 13.
400x magnification of summer larva gut epithel incubated with anti-AFP and stained with Alexa Fluor 488. Panels show fluorescence at FITC (A), DAPI (B) and TRITC (C) wavelengths (See appendix 6, 24112011 2if). High resolution FITC in Appendix 2, fig A7.

The strong, spotty fluorescence in the gut epithelium and -contents are likely to be from antibodies, but autofluorescence cannot be ruled out completely.

# Discussion

The discussion is divided into three parts. The most pressing matter is off course the significance of the results versus the reliability of the method.
Accepting the results leads into the actual discussion of localisation of AFP in Rhagium mordax larvae during winter and summer. Lastly, the results are put into a environmental and evolutionary perspective.

## Significant fluorescence

Using a qualitative method such as immunofluorescence has it's advantages, as the results can be aesthetically pleasing and easily communicated to the layman. The main disadvantage is lack of proper protocols for automation, reproducibility and statistic calculations. Such practices do exist, but the very nature of the method prevents them from being meaningfully and objectively employed in every case. Concrete evidence that the results are reliable are hard to come by in this type of work. Nevertheless, immunofluorescence may predict or verify the outcome of other methods performed on the same subject. Results of immunofluorescence studies on their own should be regarded as a tendency towards a theory, and only as evidence when in conjunction with quantitative methods. Immunofluorescence may provide functional insights that gel electrophoresis or chromatography cannot and therefore serves as a powerful tool for hypothesis or subhypothesis testing.
Based on the knowledge that larvae of the beetle Rhagium mordax are resistant to cold and contain AFP in their haemolymph, the aim of this project has been providing evidence of AFP localisation within the tissue of larvae. Using immunofluorescence there is no way of quantifying the relative levels of AFP, and a comparison between winter- and summer larvae AFP levels does not make sense on the basis of these results.
One thing that must be commented on is the fact that more than one secondary antibody was used. The fact that Alexa Fluor 350 fails to yield any positive fluorescence when used as a secondary antibody, is a problem for all of the results gained using Alexa Fluor 488. As autofluorescence is clearly abundant in the tissue sections of Rhagium mordax, an explanation for the lack of positive fluorescence when using Alexa Fluor 350 could simply be just that. Alexa Fluor 488 has en extinction coefficient more than three times larger than Alexa Fluor 350 (Invitrogen 2011). It is very likely that the signal from Alexa Fluor 350 drowned in the background autofluorescence. Because of this, the discussion will only take Alexa Fluor 488 results into consideration.
The negative controls provided for larval cuticle (fig. 8) provide a strong basis for further discussion of significant results, as they are almost indistinguishable and show strong autofluorescence in the epicuticle. The waxy layer of the cuticle in larvae of *Dendroides canadensis* autofluoresce less during summer (Olsen *et al.* 1998). The method by which brightness levels of photographs captured for this project is controlled is based on the brightest parts of the entire scene. Bright autofluorescence in the epicuticle may elevate the minimum brightness cutoff to a level above that of the epidermis, exo- endocuticle. This may help explain some of the discrepancies between negative controls (fig. 8) and the autofluorescence of summer larvae cuticle (fig. 12).
The negative controls provided as the basis of gut epithelium (fig. 9) AFP localisation are somewhat weaker. The control provides a view of the smooth autofluorescence seen in the gut epithelium aswell as the gut contents. The autofluorescence is smooth, but it is not completely identical at the three wavelengths. On the background of this autofluorescence it will be difficult to distinguish between positive fluorescence and autofluorescence. Positive fluorescence in the gut comes down to a subjective descision to a much higher degree, than the significance of fluorescence in the cuticle. The autofluorescence in the gut at FITC and TRITC is, however, a near match and the addition of bright fluorophores may tip the scale in favor of significant results.
The summer larvae were collected prior to the release of temperature data for May 2011. Even though it felt safe to collect larvae that had no chance of experiencing sub-zero temperatures, the temperature data unfortunately shows that freezing temperatures were observed on two occasions. From this knowledge, the original hypothesis could very well be changed to include AFP activity in larvae collected in May.

# Protein localisation

From previous studies of protection against inoculative freezing, it is known that insect cuticle (Olsen *et al.* 1998) and the skin of fish (Low *et al.* 1998) are very important ice propagation barriers. Therefore, the outer cuticle of Rhagium mordax larvae has been of high interest.
*Cuticle*
Winter larvae show a great deal of positive fluorescence beneath the epicuticle (fig. 10). The epidermis, endo- and exocuticle show positive fluorescence, which fits well with previous studies (Olsen *et al.* 1998, Xu & Duman 1991) that mark the epidermis as AFP producing tissue. The same holds for the cuticle of summer larvae (fig. 12).
The background autofluorescence is somewhat more significant in the winter larvae, which could be an artifact of the image capture method or as a result of altered composition of the waxy cuticle layer. The working hypothesis for this project did not include AFP in the cuticle of larvae collected during summer, due to the seemingly low usefulness of AFP during warm months. A study of the seasonal variation of AFP in Rhagium mordax shows that the haemolymph level of AFP never drops to zero (Wilkens & Ramløv 2008). It does, however, look as if much of the fluorescence in summer larvae cuticle (fig. 12) is located in the epidermis, as opposed to the winter larvae cuticle (fig. 10) where there endo- and exocuticle shine the brightest. This could off course also be specific to the tissue sections used.
The summer larvae showed a remarkably tight pattern of small flourescent spheres in the cuticle (fig. 12). This sort of extremely localized fluorescence was not observed in winter larvae and was only present in summer larvae.
*Gut*
The gut epithelium of winter larvae (fig. 11) shows positive fluorescence. Even though the negative control (fig. 9) shows the same abundant fluorescence at first glance, a closeup (Appendix 2, fig. A5) of the winter larvae gut reveals a pattern typical for antibodies. A large degree of positive fluorescence is also observed inside the gut lumen. This does not seem to be especially localised, but very smooth. Almost like autofluorescence, but probably not since the lumen is dark at DAPI and TRITC (fig. 11 B, C). The gut epithelium and -lumen of summer larvae (fig. 13) show positive fluorescence. The lumen of summer larvae is interesting as the fluorescence is localised to small circles. This is much like the pattern seen in the cuticle of summer larvae (fig. 12).
Although there are no previous studies into the exsistence of vesicle-borne AFP in insects, the results presented within this project could be the basis of a new hypothesis.

Because the mean temperature in May 2011 was close to 10°C, having a constant high level of AFP could seem unneeded or disadvantageous. Instead of a organism-wide breakdown and recycling of AFP, it now seems likely that AFPs are instead stored in vesicles near the tissue that express them during cold months. Much to the authors surprise, freezing temperatures were observed on two occasions during May. As it turns out, having AFPs in vesicles near the tissue that protects against inoculative freezing is maybe not as redundant as it looks.
A previous study in the organismal distribution of AFP isoforms in *Dendroids canadensis* showed that larvae from the same natural population could be split into sub populations. Based on the AFP isomer composition (Duman *et al.* 2002). This could of course be the case in *Rhagium mordax aswell.* A quantitative proteome analysis should be applied in a future to prove or disprove this hypothesis.

Even when taking all the caveats of the method into account, the clear positive fluorescence of the cuticle marks the location of antifreeze proteins beyond reasonable doubt.
AFP located in and around the gut lumen are a topic of discussion, but the results presented within this project point towards the gut epithelium and cuticle being equally important in preventing inoculative freezing.

# Environmental perspective

Larvae of Rhagium mordax have been known to empty their gut when the temperate winter approaches (Wilkens & Ramløv 2008). This phenomenon was not observed in larvae collected in march (the winter larvae). It is likely that their guts were empty during December, January and Febuary, but they were definitely not in march. Figure 7 shows a HE stain of a larva collected in march, with a gut lumen that is clearly not empty. AFPs were observed inside the gut lumen in both winter and summer larvae. The pattern, however, was slightly different in the two seasons, with summer larvae leaning towards spheres or vesicles of fluorescence. The lumen of winter larvae was seemingly filled with AFPs, while the summer larvae showed a somewhat sparse distribution. It looks like AFPs are abundant in the gut epithelium and actively secreted into the lumen of the gut. This mechanism could be and adaptation to enable feeding and digestion during months of moderate cold. Unless there is a mechanism for reabsorbtion of AFPs in the lumen, this practice could be very expensive in terms of energy and amino acids. This expense should off course be weighed against the ecological advantage and energy gain associated with being able to feed a few months more every year.
It is interesting to see such abundance of AFPs in the summer larvae. One could speculate that it would be more favorable to recycle the AFPs in months of moderate mean temperature (>5°C), in favor of growth. Living primarily beneath a layer of bark, the environmental temperature experienced by Rhagium mordax larvae is somewhat buffered. A study of larvae from another AFP producing insect, *Choristoneura fumiferana*, shows that reaching maximum levels of AFP requires close to two weeks of synthesis (Qin *et al.* 2007). The activity of antifreeze proteins is not strictly correlated with concentration in a linear fashion (Zachariassen *et al.* 2002). There seems to be a minimum, or threshold, concentration that must be upheld in order for AFPs to work. This is probably a function of their mode of action, as too much spacing between adsorbed AFPs leads to larger ice surfaces. Only on small ice appendages can the Kelvin effect inhibit ice growth.

With this in mind, it makes sense that larvae are reluctant of recycling antifreeze protein, even at the end of spring. This may also explain why a previous study in the haemolymph concentration of AFP showed low levels of expression all year round, with the highest concentrations observed in January (Wilkens & Ramløv 2008).
Judging by the results presented in this project, the seasonal abundance scheme observed in the haemolymph also applies to tissue protecting against inoculative freezing.

## Future expansions

To cover some of the questions raised by this study, a few ideas for future experimental work is suggested.
It was established that autofluorescence is quite formidable in tissue sections of Rhagium mordax. So much that fluorescent dyes with extinction coefficients under 20k are likely to drown in autofluorescence. Even though the DAPI autofluorescence tends to be smooth and somewhat dampened compared to FITC and TRITC, fluorescent DAPI dyes are often smaller molecules with low light absorption and emission. For this type of tissue, they do not appear to be useful. TRITC autofluorescence in Rhagium mordax is bright and contrastive. This does not make TRITC mimicking dyes an obvious choice, but on the other hand they tend to be larger molecules with higher extinction coefficients. There is the possibility that they can outshine the background fluorescence to a much higher degree than Alexa Fluor 350 and -488.
To elucidate the question of AFP localisation as a function of season, a data intensive study with bimonthly collection and visualisation of larvae could be initiated. Using more or less the same methods used in this project for immunofluorescence, but with the addition of a quantitative analysis (Wilkens & Ramløv 2008, Duman *et al.* 2002). Using brighter fluorescent dyes, AFP localisation could go beyond just cuticle and gut epithelium.
The third, and probably most important, matter is that of quantifying results. The way microscopy data is handled should be strictly systematic, and a method for calculating and normalising that data should be developed. The best suggestion is a closer familiarisation with the image capture software used. High dynamic range (HDR) capture is a technique that would fit the purpose of this very well, as light levels can be wildly different between wavelengths. The method of auto-exposure used in this project is suited for initial studies, but a more systematic approach with constant exposure and higher dynamic range in the image capture is probably better. This would be an important first step towards doing statistic calculations on immunofluorescence data.

# Conclusion

Using immunofluorescence it was observed that larvae of Rhagium mordax have antifreeze proteins in their cuticle, gut epithelium and -lumen during winter (March) and summer (May) months. Larvae secreted AFP into the gut lumen, probably an adaption to food uptake during cold months. During summer, AFPs were localised to small spheres or vesicles. Results point towards prevention of inoculative freezing as being the most important function of AFP.

# Acknowledgements

I must acknowledge the help and guidance of Professor Hans Ramløv. Primarily for letting me work independently while providing materials and equipment used in this study. Without the help of Faredin Aliyevski I would have never found any larvae in the forest, making this project quite dull. Lastly I must thank my fellow student Asbjørn Wejdling. Without his tireless commenting, proofreading and good spirits, it would not have been possible to finish in time.

# Appendix 1 - Step-by-step protocol

## Tissue sectioning

This method is a slightly modified version of one I have previously compiled (Johnsen 2011)

**1: Dehydration of fixated tissue**

Larva is removed from FineFix and cut twice perpendicular to the bilateral axis using a scalpel. As the fixated larvae can be very soft, it is important to slice and not push the scalpel through.

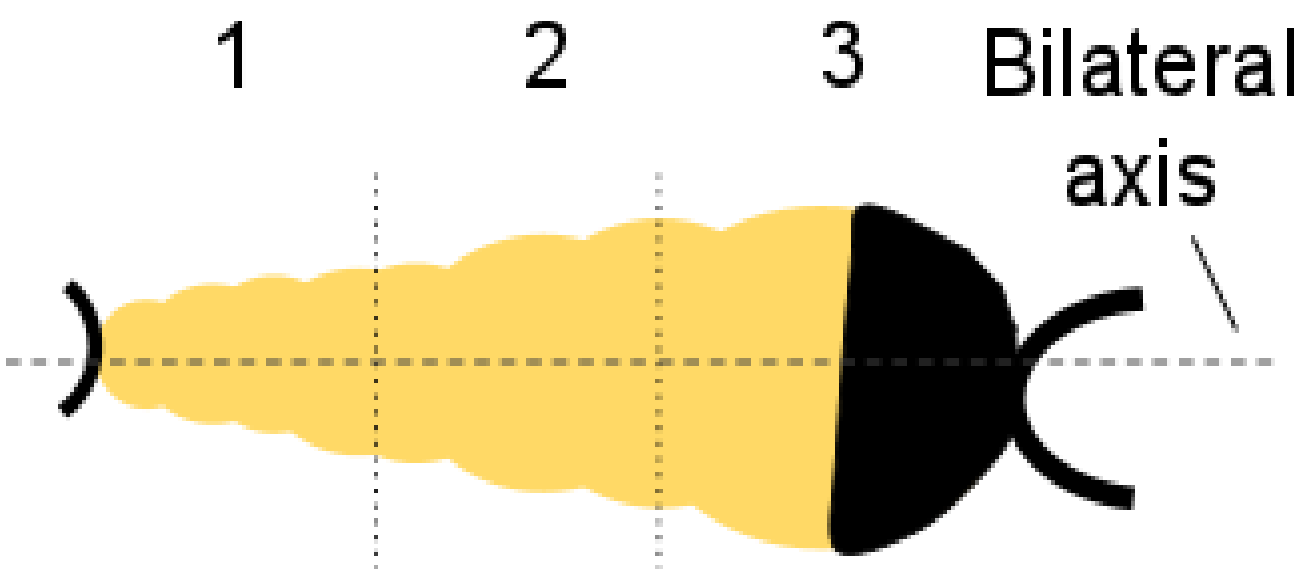

Figure A1.

The three pieces of larva are transferred to increasingly concentrated solutions of ethanol while being gently shaken (IKA Labortechnik KS 125 B S1 Orbital Shaker).

- 80% ethanol, one hour
- 95% ethanol, two times, one hour each
- 100% ethanol, three times, one hour each

Exces ethanol is removed with xylene:

Xylene is toxic and should be handled under fume hood.

- Pure xylene, three times, one hour each

**2: Embedding in paraffin wax**

Tissue is infiltrated by molten paraffin wax at ~60°C. This can be done in an embedding cassette while placed in the embedding mold (Thermo Scientific 6401016), but the cassette must be removed before the last infiltration as this is the paraffin that is allowed to harden. The paraffin wax is kept warm at the bottom of a Thermolyne Type 16500 Dri-Bath covered with aluminium foil and styrofoam.

- Molten paraffin wax, 90 minutes

Tissue sample is moved to a new mold with molten wax and left on the heater over night to ensure 100% infiltration.

- Molten paraffin wax, over night (>12 hours)

The mold is filled with clean paraffin wax one last time, and allowed to solidify. The block of paraffin and tissue is then ready for sectioning on the microtome. Note that the paraffin wax should not be allowed to harden slowly, as crystals may form. Room temperature around 20°C should be sufficiently cool.

- Remove tissue from cassette and place in base mold filled with molten paraffin.

Wait for the tissue to settle and the paraffin wax to become completely fluid again. The best way to position the larval tissue is by having the bilateral axis perpendicular to the bottom of the mold. This way, every section will contain more or less the same tissue and be easier to compare.

- Remove mold from heater and allow the wax to harden.

**3: Cutting and mounting**

Mount paraffin block in Biocut 2030 microtome and cut sections of 5 µm at a 5° angle.

Sections are transferred to a flotation bath (Agar Scientific, L4136) filled with deionized water at 45-50 °C using metal instruments coated with paraffin.

Expanded sections are quickly and carefully collected directly onto polysine slides (Agar Scientific L4345) from the flotation bath.

Slides are dried in an oven set to 40°C for at least two hours.

**4: Removal of paraffin and rehydration**

Slides are placed in a slide staining holder (Bio Optica, 10-42) for faster processing. Each step is carried out in a separate slide staining dish (Bio Optica, 10-30).

- Slides are incubated in xylene for 10 minutes to remove paraffin wax.

Rehydration of tissue is done using decreasing concentrations of ethanol:

- 100% ethanol, two minutes
- 95% ethanol, one minute

Slides are rehydrated with distilled water

- Distilled water, one minute.

From now on, the slides should not be allowed to dehydrate, as it can obscure antigens.

The tissue sections are now ready for immunostaining or HE staining.

# Immuno staining

**1: Blocking and staining**

- Mark area around section with PAP pen (Agar Scientific, L4197S) to avoid runoff of antibodies or blocking solution.

- Place slides in steel incubator (Agar Scientific, L4474).

- Add 100 µl of blocking solution per slide and incubate at room temperature for 30 minutes with the lid closed.

Drain blocking solution from slides by tilting and gently touching the surface of the water with lens paper.

- Add 100 µl of primary antibody per slide. Incubate at 4°C overnight.

- Wash slides in 1x TBS 4 times, 5 minutes each.

- Rinse once with 5% BSA in TBS, 5 minutes.

Secondary antibody contains a photosensitive region, so do the following steps should be performed in the dark, if possible.

- Add 100 µl secondary antibody and incubate at room temperature for 30 minutes.

- Wash with 1x TBS 4 times, 5 minutes each.

**2: Mounting and visualization**

Slides are mounted and coverslipped using a drop Citifluor AF1 (Agar Scientific, R1323).

Slides are visualized on an Axio Imager.M2m setup. Only FITC fluorescence that does not overlap with TRITC or DAPI fluorescence should be considered a significant result.

# HE staining

The series of incubations involved in performing an HE stain should be prepared in individual slide staining dishes before the first incubation with xylene is initiated. Slides should be processed in batches using a slide staining holder. Note that this method is based on a modified version of the one described in the instructions manual provided with Shandon Instant Hematoxylin.

1. Five minutes in hematoxylin
2. One minute in distilled water

3. 15 seconds in acidic alcohol
4. One minute in distilled water
5. One minute in bluing reagent
6. One minute in distilled water
7. One minute in 80% ethanol
8. 30 seconds in alcoholic eosin
9. One minute in 95% ethanol
10. Two minutes in 100% ethanol
11. Five minutes in xylene

Excess xylene should be allowed to evaporate in a fume hood for a few minutes before mounting and coverslipping.

Slides were mounted using 100 µl Shandon ClearVue XYL (Thermo Scientific 4212) per slide.

# Appendix 2 - High resolution immunofluorescence

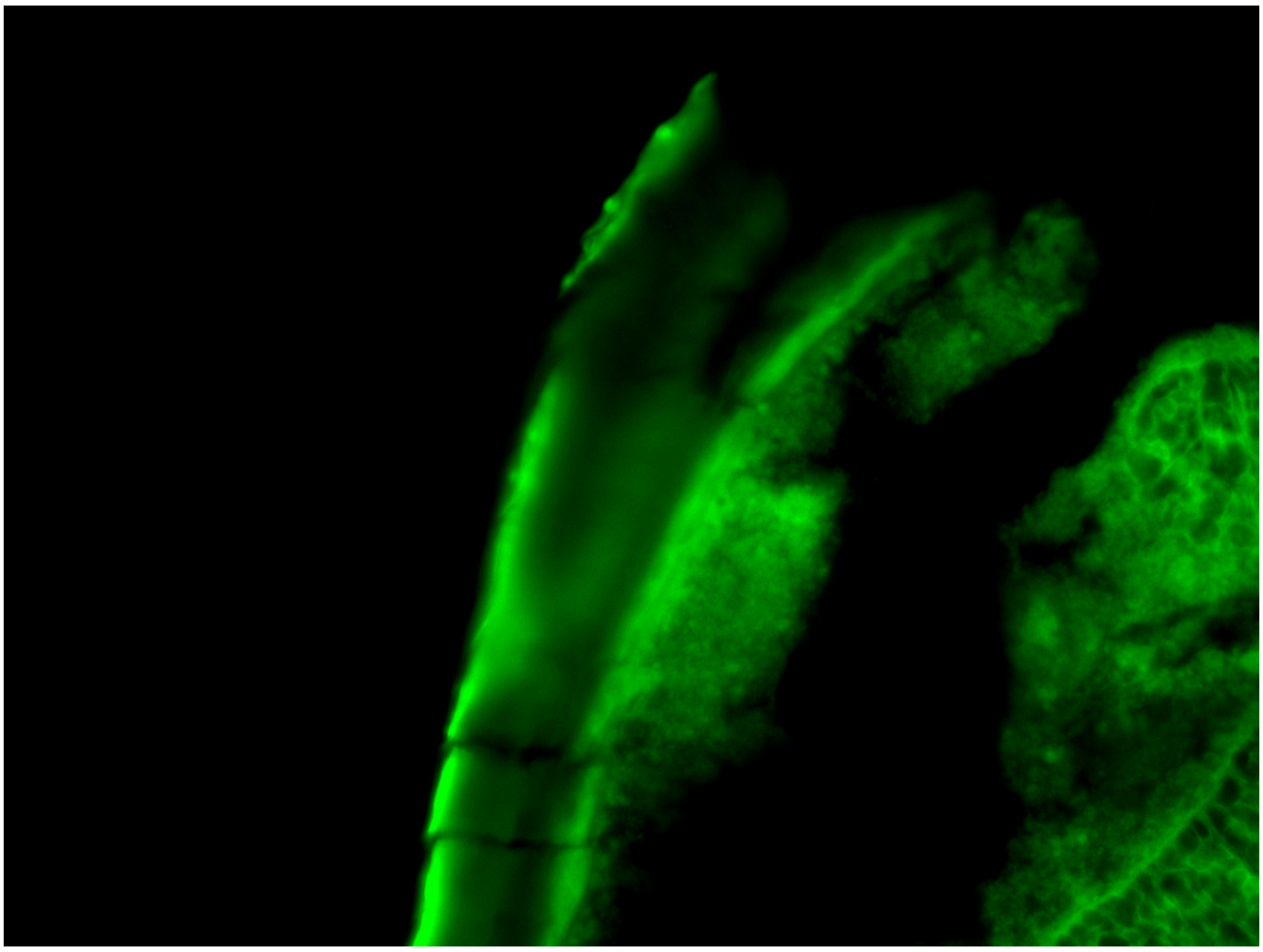

Figure A2 – High resolution autofluorescence in winter larva cuticle.

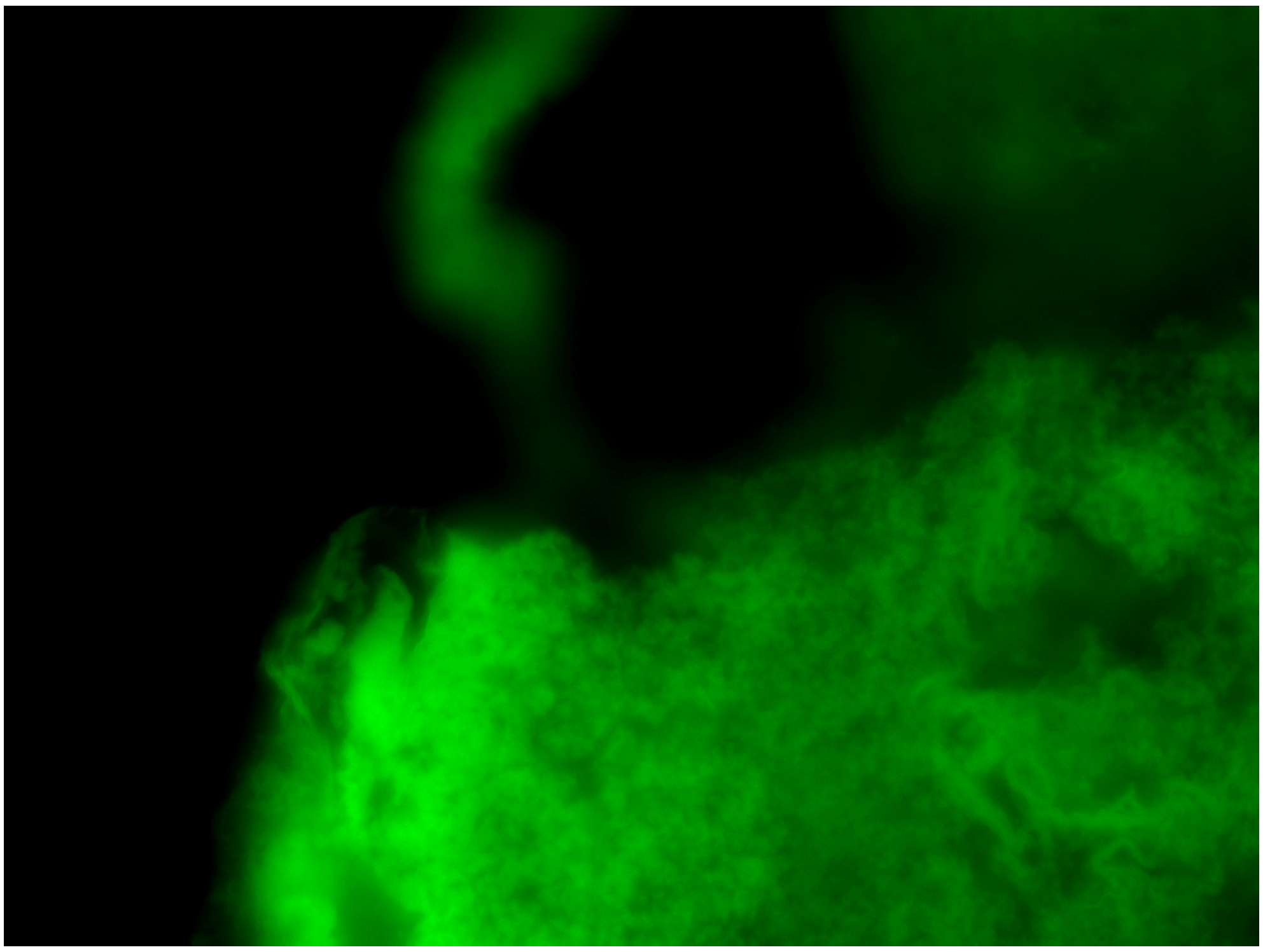

Figure A3 - High resolution autofluorescence in winter larva gut.

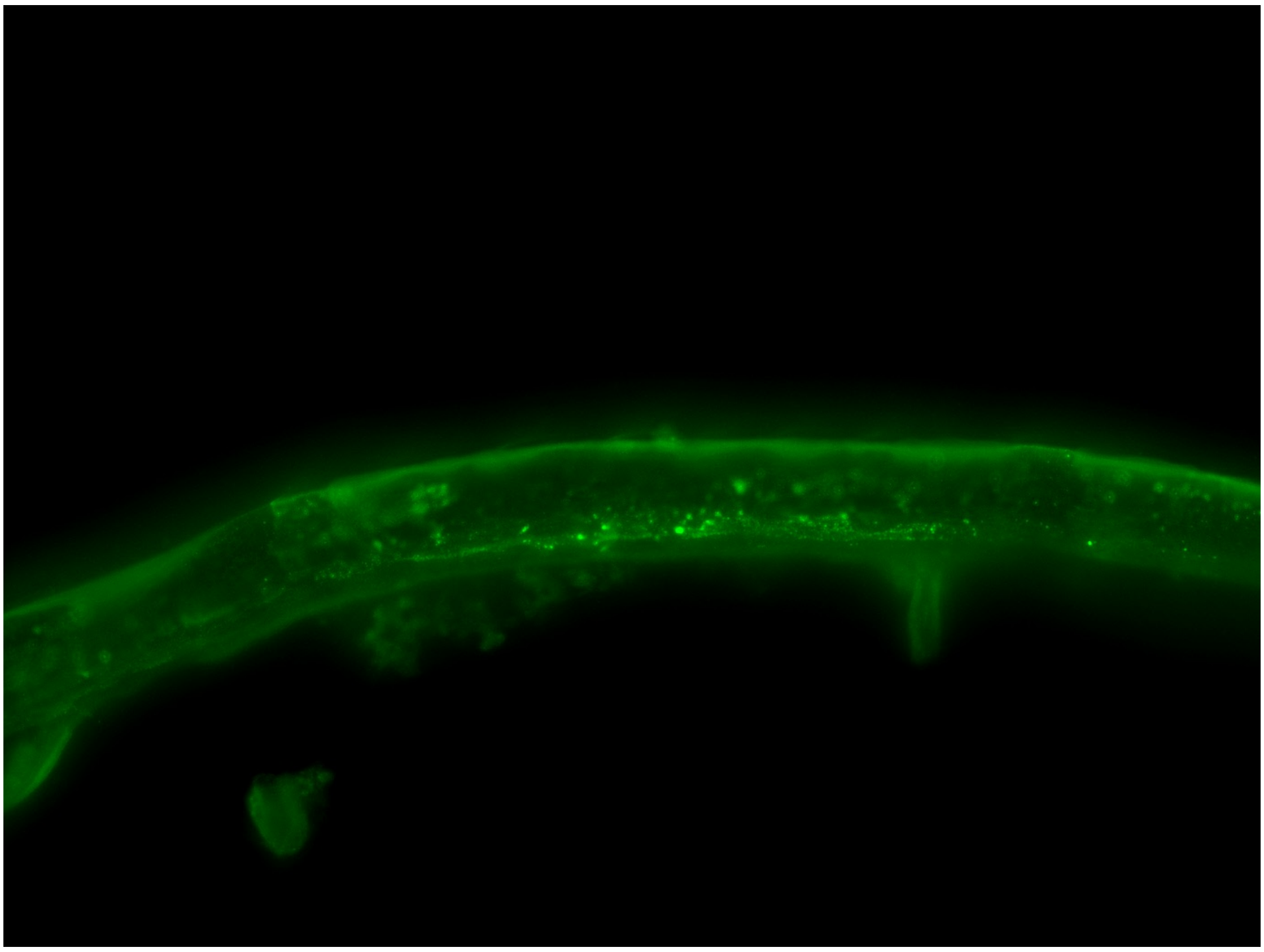
Figure A4 - High resolution immunofluorescence in winter larva cuticle.

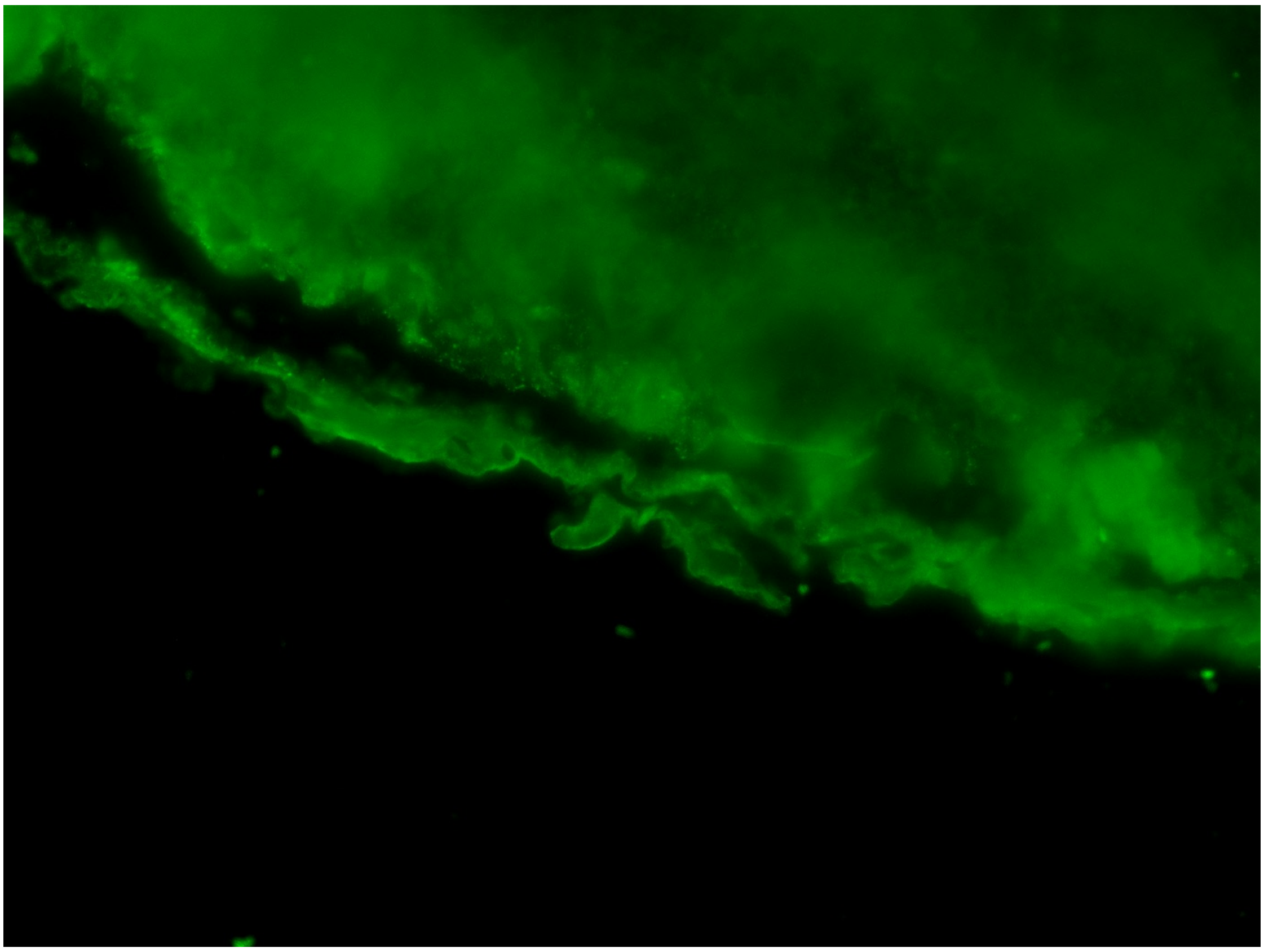
Figure A5 - High resolution immunofluorescence in winter larva gut.

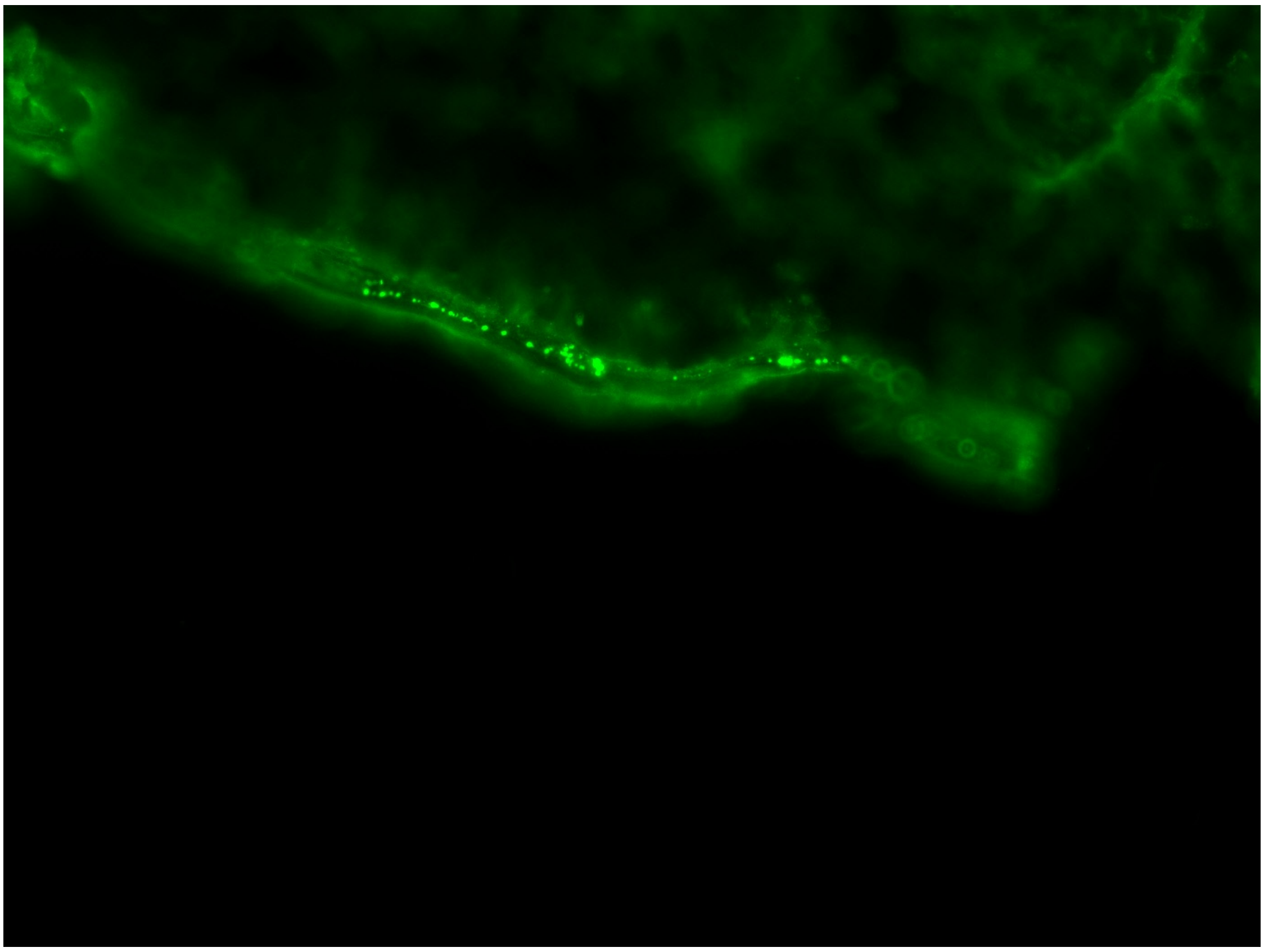

Figure A6 - High resolution immunofluorescence in summer larva cuticle.

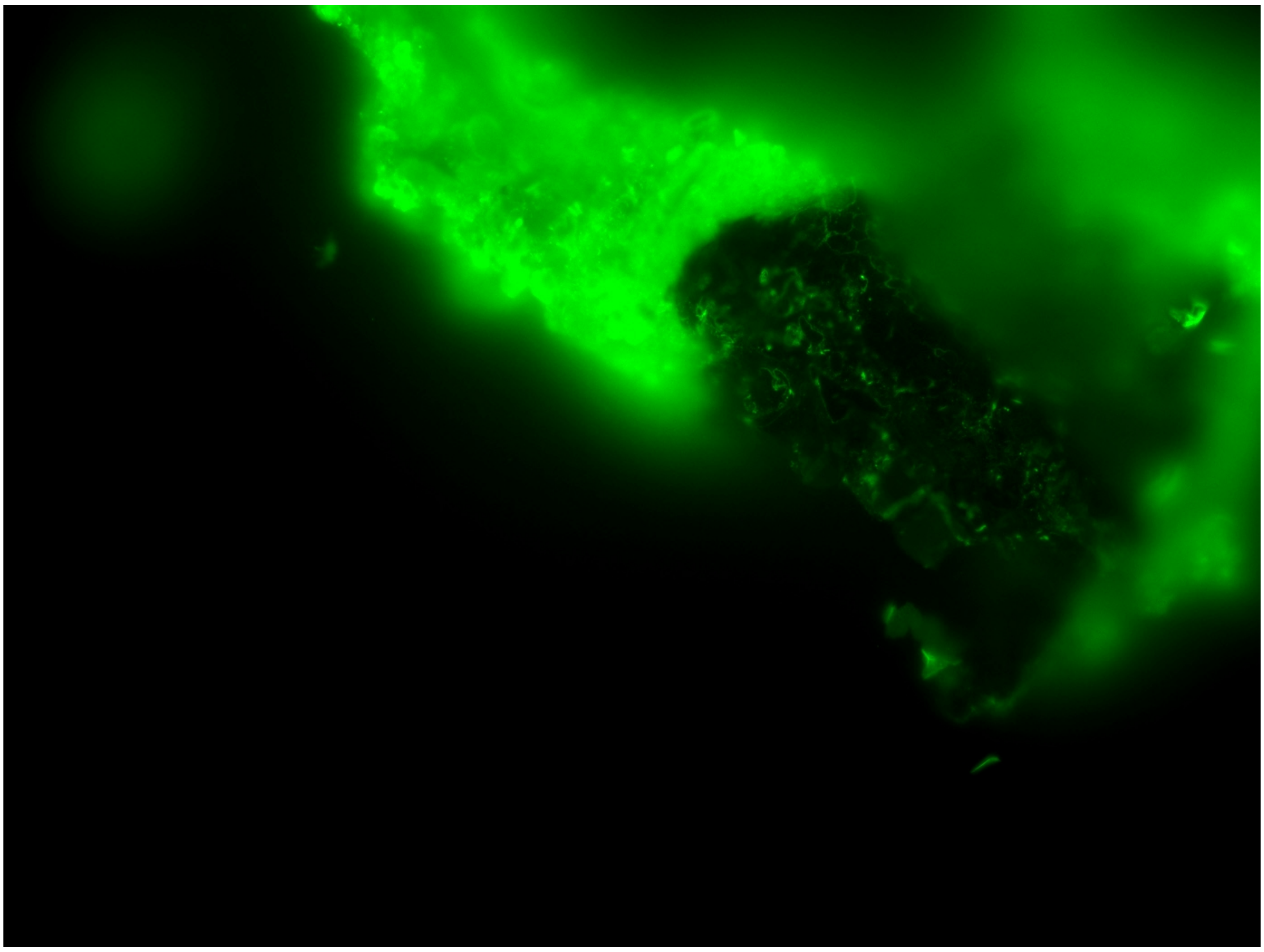

Figure A7 - High resolution immunofluorescence in winter larva gut.

# Appendix 3 - Sequence alignment

Sequence alignment using http://xylian.igh.cnrs.fr/bin/align-guess.cgi (Pearson *et al.* 1997)

```
>_ riafp                                         134 aa vs.
>_ rmafp                                         135 aa
scoring matrix: , gap penalties: -12/-2
75.0% identity;         Global alignment score: 596

                 10        20        30        40        50
170828 --CRAVGVDGRAVTDIQGTCHAKATGAGAMASGTSEPGSTSTATATGRGATARSTSTGRG
         :::::::. .:::.:::::::::: ::.:::::  ::::::::: :::: ::::: :
_      YSCRAVGVDASTVTDVQGTCHAKATGPGAVASGTSVDGSTSTATATGSGATATSTSTGTG
               10        20        30        40        50        60

       60        70        80        90       100       110
170828 TATTTATGTASATSNAIGQGTATTTATGSAGGRATGSATTSSSASQPTQTQTITGPGFQT
       :::::::..:.::::::::::::.::::.:..:: ::.:::.::..::::.:..::: ::
_      TATTTATSNAAATSNAIGQGTATSTATGTAAARAIGSSTTSASATEPTQTKTVSGPG-QT
               70        80        90       100       110

      120       130
170828 AKSFARNTATTTVTAS
       : ..: .:::::::::
_      ATAIAIDTATTTVTAS
     120       130
```

# Appendix 4 - Fluorophores of interest

All information gathered on http://products.invitrogen.com (Invitrogen 2011)

| Name | Absorption maximum | Emission maximum |
|---|---|---|
| FITC | 494 nm | 518 nm |
| DAPI | 358 nm | 461 nm |
| TRITC | 555 nm | 580 nm |
| Alexa fluor 350 | 346 nm | 445 nm |
| Alexa fluor 488 | 494 nm | 519 nm |

# Appendix 5 - Solutions

## Common components

Xylene (Merck 8086972500)
Ethanol (Kemetyl 96% finsprit)
37% HCl (Merck 1003172500)
Shandon histoplast / paraffin wax (Thermo Scientific 6774006)

**FineFix working solution**

720 ml ethanol
280 ml FineFix concentrate (HD Scientific ML 70147)

## HE stain

**Bluing reagent (0.1% Sodium bicarbonate)**

1 g Sodium hydrogen carbonate (Merck 1063232500)
1000 ml distilled water
Mix to dissolve.

**Acidic alcohol (0.3% HCl in dilute ethanol)**

280 ml 96% ethanol
120 ml distilled water

1.2 ml 37% HCl
Mix and store in sealed container.

**Hematoxylin**

Shandon Instant Hematoxylin (Thermo Scientific 6765015)
Prepared from two component dry powder as per included instructions.

**Eosine**

Wright's eosin methylene blue solution (Merck 101383)

# Immuno stain

**10x TBS (0.5M Tris Base, 9% NaCl, pH 8.4)**

61 g Trizma base (Sigma T1503)
90 g NaCl (AppliChem 3597)
1 L distilled water
Mix to dissolve. Adjust pH to 8.4 using fuming HCl. Dilute 1:10 before using as wash buffer
or solvent for blocking solution.

**Blocking solution (5% BSA in TBS)**

10g BSA powder (Sigma A4503)
200 ml 1x TBS
Mix to dissolve. Note that it may take some time, but the BSA will eventually dissolve.

## Primary antibody

Primary antiserum (BioGenes, Product no. 41117, animal no. 20140, RmAFP#1)
Primary antibody was diluted by one addition of:
20 µl serum + 980 µl blocking solution = 1:50

## Secondary antibody

Secondary antibody stock (Invitrogen A11070)
Secondary antibody was diluted using a 1:100 -> 1:1000 successive dilution.
1 µl (2mg/ml Anti-rabbit IgG) + 99 µl blocking solution = 1:100
10 µl (1:100) + 990 µl blocking solution = 1:1000

This corresponds to the following operational concentrations for secondary antibodies:

| Name | Concentration |
|---|---|
| Stock | 2 mg/ml |
| 1:100 | 20 µg/ml |
| 1:1000 | 2 µg/ml |

# Appendix 6 - Overview of experiments

| Series (date) | # | IF/HE | Primary dilution factor /BSA | Secondary dilution factor / Alexa Fluor |
|---|---|---|---|---|
| 11112011 | 1 | HE | | |
| 11112011 | 2 | HE | | |
| 11112011 | 3 | IF | 100 | 1000 / 350 |
| 11112011 | 4 | IF | 200 | 1000 / 350 |
| 11112011 | 5 | IF | 100 | 2000 / 350 |
| 11112011 | 6 | IF | 200 | 2000 / 350 |
| 11112011 | 7 | IF | BSA | 1000 / 350 |
| 11112011 | 8 | IF | BSA | 2000 / 350 |

Pilot study that showed the lack of positive fluorescence in Alexa Fluor 350. Winter larva.

| | | | | |
|---|---|---|---|---|
| 15112011 | 1 | HE | | |
| 15112011 | 2 | IF | 1 | 10 / 488 |
| 15112011 | 3 | IF | 50 | 100 / 350 |
| 15112011 | 4 | IF | 50 | 100 / 488 |
| 15112011 | 5 | IF | 50 | 1000 / 350 |
| 15112011 | 6 | IF | 50 | 1000 / 488 |
| 15112011 | 7 | IF | BSA | 100 / 350 |
| 15112011 | 8 | IF | BSA | 100 / 488 |
| 15112011 | 9 | IF | 50 | BSA |
| 15112011 | 10 | IF | 50 | BSA |
| 15112011 | 11 | IF | BSA | BSA |

Pilot study that showed the combination of 1:50 + 1:1000 / 488 working well. Winter Larva.

| | | | | |
|---|---|---|---|---|
| 18112011 | 1 | HE | | |
| 18112011 | 2 | IF | 50 | 1000 / 350 |
| 18112011 | 3 | IF | 50 | 1000 / 350 |
| 18112011 | 4 | IF | 50 | 1000 / 350 |
| 18112011 | 5 | IF | 50 | 1000 / 488 |
| 18112011 | 6 | IF | 50 | 1000 / 488 |
| 18112011 | 7 | IF | 50 | 1000 / 488 |
| 18112011 | 8 | IF | 50 | 1000 / 350 |
| 18112011 | 9 | IF | 50 | 1000 / 488 |
| 18112011 | 10 | IF | BSA | BSA |

First real experiment with winter larvae.

| | | | | |
|---|---|---|---|---|
| 22112011 | 1 | IF | 50 | 1000 / 488 |
| 22112011 | 2 | IF | 50 | 1000 / 488 |
| 22112011 | 3 | IF | BSA | 1000 / 488 |
| 22112011 | 4 | IF | BSA | 1000 / 488 |

Using the head-piece from the 18112011 larvae to perform missing controls from 18112011

| | | | | |
|---|---|---|---|---|
| 24112011 | 1 | IF | 50 | 1000 / 488 |
| 24112011 | 2 | IF | 50 | 1000 / 488 |
| 24112011 | 3 | IF | BSA | 1000 / 488 |
| 24112011 | 4 | IF | BSA | 1000 / 488 |

First summer larva. Paraffin was too cold when sectioned. Samples were not very good. Lack of slides prevented DAPI samples.

| | | | | |
|---|---|---|---|---|
| 30112011 | 1 | IF | 50 | 1000 / 488 |
| 30112011 | 2 | IF | 50 | 1000 / 350 |
| 30112011 | 3 | IF | 50 | 1000 / 488 |
| 30112011 | 4 | IF | 50 | 1000 / 350 |
| 30112011 | 5 | IF | BSA | 1000 / 488 |
| 30112011 | 6 | IF | BSA | 1000 / 350 |
| 30112011 | 7 | HE | | |

Second summer larva. Paraffin was sectioned at correct temperature. New slides available, DAPI samples made.

# Appendix 7 - Alexa Fluor 350

The immunofluorescence results from sections stained with Alexa Fluor 350 instead of Alexa Fluor 488 were meant as a control of the -488 experiments. It turned out that the extinction factor of the 350 variety was simple too low, relative to the autofluorescent background.

# Alexa Fluor 350 fluorescence

Alexa Fluor 350 did not result in any positive fluorescence. There may be weak signs of significance, but it is very circumstancial. This appendix is structured like Alexa Fluor 488 results.

## Autofluorescence and unspecific binding

As with the Alexa Fluor 488 results, unspecific binding and autofluorescence will be first. The two controls are made with a summer larva, unlike the winter larva used for Alexa Fluor 488 controls. AFP are plentiful in summer larvae, so this should not cause any problems later when interpreting the results.

### Cuticle autofluorescence

Autofluorescence is clearly visible in the epicuticle of summer larvae. The autofluorescence in this layer is strong, but unlike winter larvae the underlying layers are vaguely visible. There is a slight background at DAPI (A2A), FITC (A2B) and TRITC (fig. A2C).

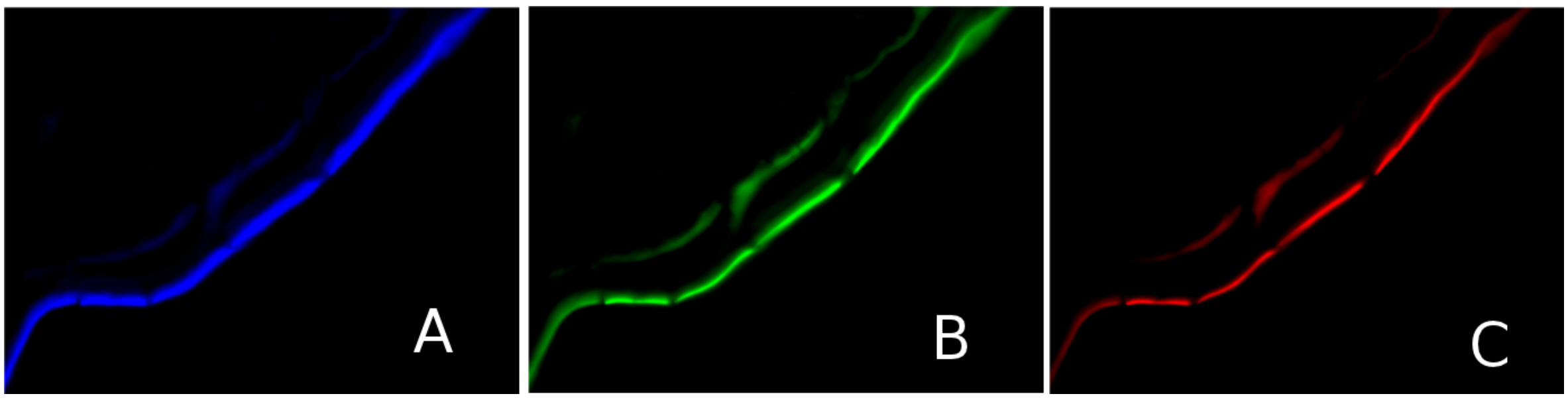

Figure A2
400x magnification of summer larva cuticle incubated with no primary antibody and stained with Alexa Fluor 350. Panels show fluorescence at DAPI (A), FITC (B) and TRITC (C) wavelengths.
(Appendix 6, 30112011 6if)

Based on my negative control (fig. A2), it seems plausible that any cuticular fluorescence other than that in the epicutle, can be regarded as positive fluorescence, if antibodies can shine through the autofluorescence.

### Gut epithelium autofluorescence

Unlike the uneven autofluourescence in the cuticle, the area around the lumen of the gut fluoresce more smoothly. The spongy area at the bottom is the contents of the gut, and the more brightly lit part at DAPI (fig. A3A) is the gut epithelium.

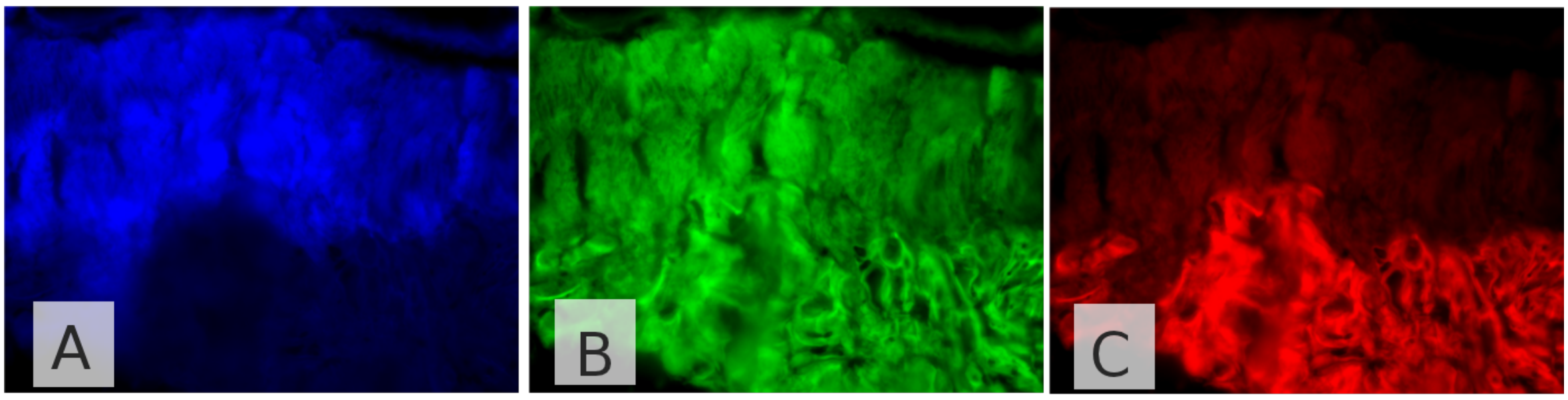

Figure A3.
400x magnification of summer larva gut epithel incubated with no primary antibody and stained with Alexa Fluor 350. Panels show fluorescence at DAPI (A), FITC (B) and TRITC (C) wavelengths.
(Appendix 6, 30112011 6if)

Autofluorescence is clearly a a widespread phenomenon in both gut and cuticle. Even if the method works succesfully, there is a risk that fluorescence from the antibodies drown in autofluorescence.

## Winter larvae

Winter larvae were stained with Alexa Fluor 350, but yielded no positive results.

### Cuticle

There is no significant difference between fluorescence at all three wavelengths. Therefore no positive fluorescence is observed.

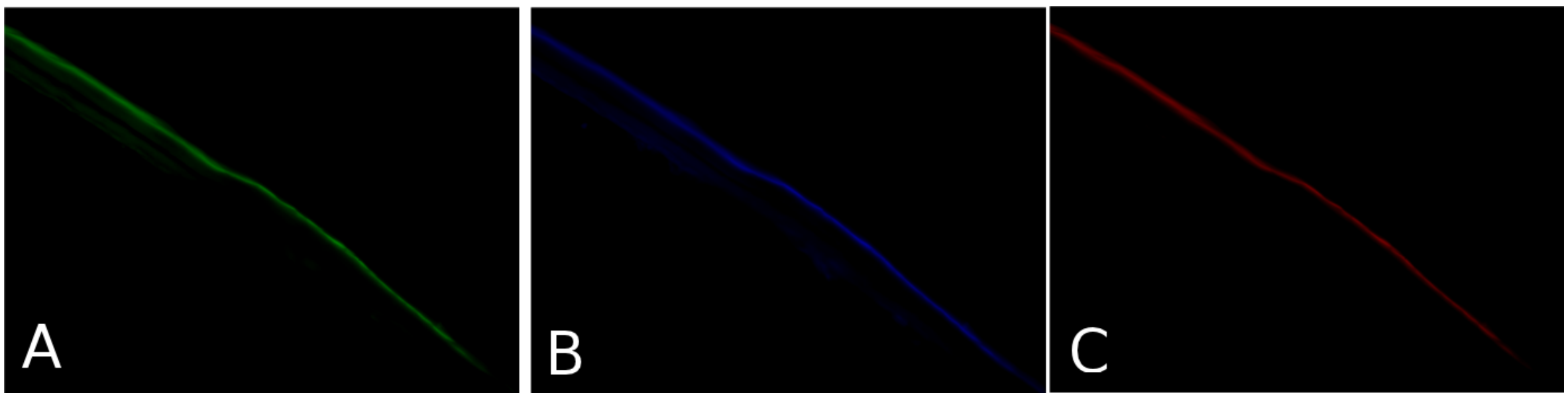

Figure A4.
400x magnification of winter larva cuticle incubated with anti-AFP and stained with Alexa Fluor 350. Panels show fluorescence at FITC (A), DAPI (B) and TRITC (C) wavelengths.
(Appendix 6, 18112011 3if)

### Gut

The same goes for the gut of winter larvae. No positive fluorescence can be observed.

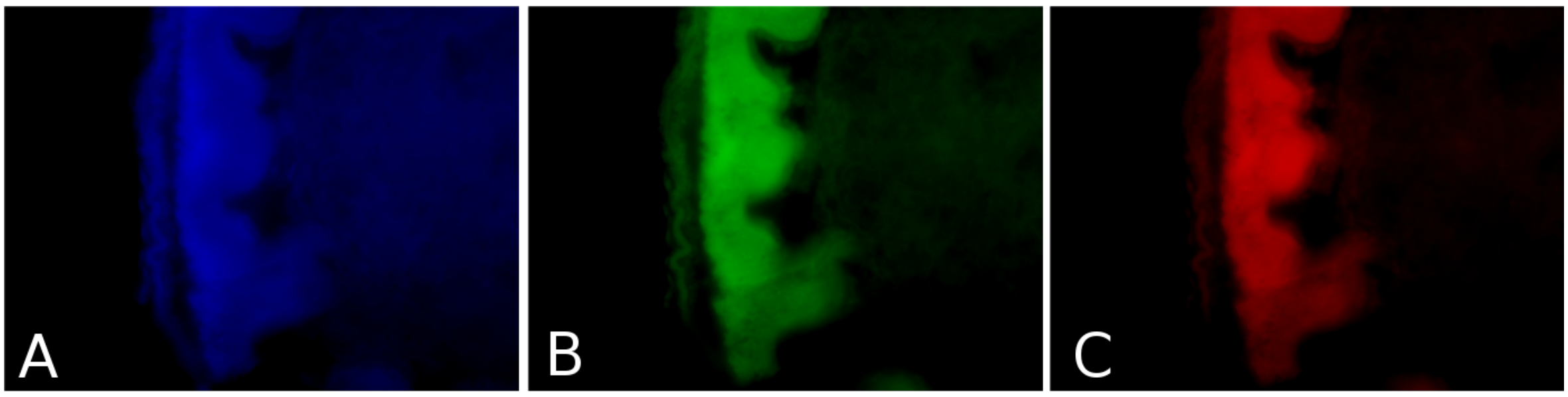

Figure A5.
400x magnification of winter larva gut epithel incubated with anti-AFP and stained with Alexa Fluor 350. Panels show fluorescence at DAPI (A), FITC (B) and TRITC (C) wavelengths.
(Appendix 6, 18112011 8if)

## Summer larvae

Summer larvae were also stained with Alexa Fluor 350, but provided only very circumstantial results. Due to the spherical grouping of AFP observed using Alexa Fluor 488, such a pattern has also been the key interest in the Alexa Fluor 350 stains.

### Cuticle

The DAPI (Fig. A6A) panel may show positive fluorescence located in small circles on the cuticle, but the result is very weak.

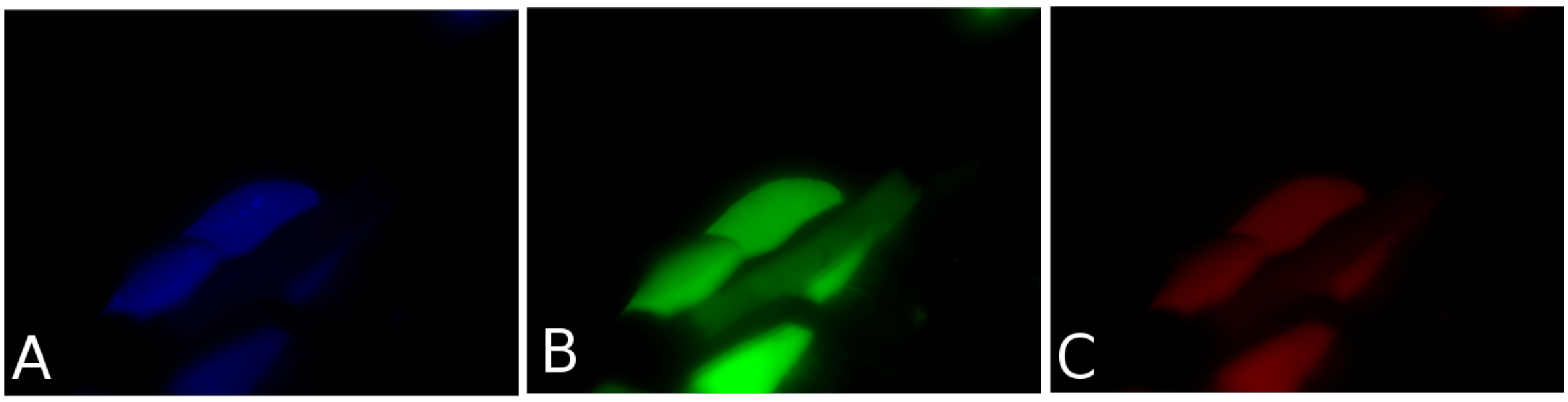

Figure A6.
630x magnification of summer larva cuticle incubated with anti-AFP and stained with Alexa Fluor 350. Panels show fluorescence at DAPI (A), FITC (B) and TRITC (C) wavelengths.
(Appendix 6, 30112011 4if)

### Gut

Due to the shifting location of autofluorescence in gut epithelium (fig. A3), no scene with the possibility of positive fluorescence could be captured. This is unfortunate, but a remedy for this was not prioritized due to the general lack of fluorescence in Alexa Fluor 350 experiments.